\documentclass{aa}  

\usepackage{graphicx}
\usepackage{txfonts}
\usepackage[usenames,dvipsnames]{color}
\usepackage[hyperfootnotes=false]{hyperref}
\hypersetup{
  colorlinks,
  citecolor=Violet,
  linkcolor=Blue, 
  urlcolor=Blue}
\newcommand{\sgn}{\text{sgn}}

\begin{document} 

   \title{Kinematic signatures of planet-disk interactions\\ in VSI-turbulent protoplanetary disks}
   \titlerunning{Kinematic signatures of planet-disk interactions in VSI-turbulent protoplanetary disks}

   \author{Marcelo Barraza-Alfaro
        \inst{1,2}
        \and
        Mario Flock\inst{1}
        \and
        Thomas Henning\inst{1}
        }
    \authorrunning{M. Barraza-Alfaro et al.}

   \institute{Max Planck Institute for Astronomy, Königstuhl 17, D-69117 Heidelberg, Germany\\
            \email{barraza@mpia.de}
        \and Department of Earth, Atmospheric, and Planetary Sciences, Massachusetts Institute of Technology, Cambridge, MA 02139, USA\\
            }

   \date{Received X; accepted Y}

 
  \abstract
   {Planets are thought to form inside weakly ionized regions of protoplanetary disks, in which turbulence creates ideal conditions for solid growth. However, the nature of this turbulence is still uncertain. In fast cooling parts of this zone the vertical shear instability (VSI) can operate, inducing a low level of gas turbulence and large-scale gas motions. Resolving kinematic signatures of active VSI could reveal the origin of turbulence in planet-forming disk regions. However, an exploration of kinematic signatures of the interplay between VSI and forming planets is needed for a correct interpretation of radio interferometric observations. A robust detection of VSI would open the door for a deeper understanding of the impact of gas turbulence on planet formation.}
   {The objective of this study is to explore the effect of the VSI on the disk substructures triggered by an embedded fairly massive planet. We will focus on the impact of this interplay on CO kinematic observations with the ALMA interferometer.}
   {We conducted global 3D hydrodynamical simulations of VSI-unstable disks with and without embedded massive planets, exploring Saturn- and Jupiter-mass cases. We studied the effect of planets on the VSI gas dynamics, comparing with viscous disks.
   Post-processing the simulations with a radiative transfer code, we examined the kinematic signatures expected in CO molecular line emission, varying disk inclination. Further, we simulate deep ALMA high-resolution observations of our synthetic images, to test the observability of VSI and planetary signatures.}
   {The embedded planet produces a damping of the VSI along a radial region, most effective at the disk midplane. For the Saturn case, the VSI modes are distorted by the planet's spirals producing mixed kinematic signatures. For the Jupiter case, the planet's influence dominates the overall disk gas kinematics.}
   {The presence of massive planets embedded in the disk can weaken the VSI large-scale gas flows, limiting its observability in CO kinematic observations with ALMA.}


   \maketitle
%

\section{Introduction}

The detection of thousands of exoplanets combined with recent observations of signatures of embedded planets in young gas-rich protoplanetary disks demonstrates that planet formation is an ubiquitous process in nature, and possibly fast and efficient. Nevertheless, the fundamental processes that pave the way from micron-sized dust particles to a planetary embryo are still far from understood. In particular, precise knowledge of the impact of turbulence on the planet formation process and disk evolution is still one of the missing links to connect the early stages of circumstellar disks to mature planetary systems \citep{Drazkowska2023}. Gas turbulence can drive angular momentum transport, and substantially affect the disk dust size and dynamical evolution, the formation of substructures, the disk thermo-chemical evolution, planetary accretion of gas and pebbles, and planet migration \citep{Lesur2023}. Therefore, understanding the origin of turbulence is crucial for further progress in planet formation theories and the interpretation of disk observations. As an example, turbulence is an essential Solar Nebula property to draw back the formation history of our Solar System \citep{Lenz2020}.

Various disk instabilities have been studied to comprehend turbulence in circumstellar disks, evoking different disk local physical conditions; therefore, possibly operating within the disk in concert. The dominant source of turbulence at each location would depend, for example, on the disk environment, disk and stellar host properties, and the specific disk region's conditions \citep{Malygin2017, Pfeil2019, Lyra2019, Lesur2023}. Among the candidates, we can summarize the magneto-rotational instability \citep[MRI, ][]{Balbus1991, Hawley1991, Balbus1998}, the vertical shear instability \citep[VSI, ][]{Nelson2013}, the convective overstability \citep[COV, ][]{Klahr2014}, and the zombie vortex instability \citep[ZVI, ][]{Marcus2015}; for a summary of these instabilities see, for example, \citet{Lesur2023} and \citet{Lyra2019}. Therefore, unraveling the physical processes behind disk turbulence requires careful analysis, comparing directly resolved observations of protoplanetary disks against state-of-the-art numerical simulations.

On constraining turbulence levels in protoplanetary disks ALMA dust and gas molecular line observations have been fundamental, providing evidence for weak turbulence in the probed disk regions \citep[e.g.,][]{Teague2016, Teague2018, Flaherty2018, Flaherty2020, Villenave2020, Villenave2022}. In these regions, tens of au from the central star, purely hydrodynamical instabilities are consistent with the expected low disk ionization \citep[i.e., MRI-dead zone,][]{Gammie1996} and the observed turbulence upper limits. Among the candidates, the VSI is the most likely to operate on these outer regions, due to the predicted fast cooling rates \citep{Malygin2017, Pfeil2019, Lyra2019}. Further, spatially and spectrally resolved CO kinematic observations have the potential to confirm such a scenario, detecting coherent corrugated flows driven by VSI \citep{Barraza2021}.

Recent ALMA molecular line observations have demonstrated the feasibility of fully resolving the gas kinematic structure of planet-forming disks, revealing the signatures of perturbations of the disk (sub-)Keplerian gas flow \citep{Teague2019, DiskDynamics2020, Teague2021, Teague2022, Garg2022, Wolfer2023, Galloway2023, Pinte2023}. These deviations from Keplerian rotation could be a dynamical manifestation of disk instabilities \citep{Hall2020, Barraza2021}, but embedded planets are often favored \citep[e.g.,][]{Izquierdo2021b, Stadler2023, Pinte2023}. In addition to substructures present in dust observations \citep[e.g.,][]{vanBoekel2017, Andrews2018, Segura-Cox2020, Garufi2018, Bae2023, Benisty2023}, this points towards an unseen population of fairly massive planets embedded in the observed protoplanetary disks \citep[see ][]{AsensioTorres2021}, with the exception of the directly observed PDS 70 b and c \citep{Keppler2018,Benisty2021}.

If massive planets are embedded in the disk (masses comparable to and above the thermal mass, see \citealt{Lin1993} and \citealt{Goodman2001}), they can strongly modify the disk density and dynamical structure \citep{Keppler2019, Bae2019, Paardekooper2023}. Interactions between massive planets and the gaseous protoplanetary disk produces a depleted gap \citep[e.g.,][]{Crida2007, Bergez2020}, spiral wakes around the planet's location \citep[e.g.,][]{Perez2018}, large-scale spiral arms via Lindblad resonances and buoyancy resonances \citep[e.g.,][]{Bae2018, Bae2018b, Bae2021}, anti-cyclonic Rossby-wave vortices \citep{Lovelace1999, Li2000, deValBorro2007}, meridional flows \citep[e.g.,][]{Kley2001, Fung2016, Teague2019c}, and a circumplanetary disk \citep[e.g.,][]{DAngelo2002, DAngelo2003, Gressel2013, Szulagyi2014}. This myriad of structures has a substantial impact on the disk gas velocities \citep[e.g.,][]{Rabago2021}. This implies that it has a direct imprint on the observed disk kinematic structure \citep{Perez2015, Perez2018, Teague2019c, Bae2021}. Moreover, planets could potentially affect the development of the VSI in the disk \citep{Stoll2017, Ziampras2023, Hammer2023}; thus, the observability of VSI signatures.

In this work, we study the kinematic structure resulting from the interplay between vertical shear instability and the structures triggered by a fairly massive planet.
The paper is structured as follows. We describe the numerical methods for the hydrodynamical simulations in Section \ref{hydrosimulations}, and present their results in \ref{simulationsresults}. The radiative transfer post-processing, simulated observations, and techniques to extract observables are detailed in Section \ref{postprocessing}, whose results are presented in Section \ref{rtresults}. In Section \ref{vsiplanetsdiscussion}, we discuss a potential approach to confirm VSI signatures observationally, and the limitations of our work. Finally, we summarize the main findings of our study in Section \ref{conclusions}.

\section{Hydrodynamical Simulations: Methods}\label{hydrosimulations}

We performed 3D global hydrodynamical simulations using the Godunov grid based code \textsc{PLUTO}\footnote{\url{http://plutocode.ph.unito.it/}} \citep{Mignone2007}. We used the publicly available version 4.4 of \textsc{PLUTO}, solving the Navier-Stokes equations of classical fluid dynamics, without magnetic fields (HD module).

\begin{equation}
    \frac{\partial \rho}{\partial t}+\vec{\nabla}\cdot (\rho\vec{v})=0
\end{equation}
\begin{equation}
    \frac{\partial (\rho \vec{v})}{\partial t}+\vec{\nabla}\cdot (\rho\, \vec{v}\, \vec{v}^\text{T})=-\vec{\nabla}P - \rho \vec{\nabla}\Phi+\vec{\nabla}\cdot \Pi, 
\end{equation}
\noindent were $\rho$ is the gas mass density, $\vec{v}$ is the gas velocity vector, $P$ is the pressure, $\Phi$ is the gravitational potential, and $\Pi$ represents the viscous stress tensor. The viscous term is included in the momentum equation only for our $\alpha$-viscous simulations, while our VSI-unstable disk simulations are inviscid. In our set of simulations, the fluid is affected by the gravitational potentials of a star ($\Phi_{\star}= -G M_{\star}/r$) and of embedded planets (see Eq. \ref{eq:planetpotential}). The disk self-gravity is not considered in our simulations.

The hydrodynamical equations were solved using a second-order accurate scheme with linear spatial reconstruction, with the least diffuse limiter implemented in \textsc{PLUTO} (monotonized central difference limiter). For the time stepping calculation, we chose the second-order Runge-Kutta time-stepping, while for the solver we use the Harten-Lax-Van Leer Riemann solver. The Courant number is set to 0.3.

We ran a total of 5 locally isothermal simulations, 3 VSI-unstable disks simulations to compare the case without planets, with a Saturn planet and  with a Jupiter planet, and 2 $\alpha$-viscous simulations of planet-disk interactions to confront the VSI-unstable cases. For all simulations, the initial conditions of the disk follow the equilibrium solutions of a disk with vertical shear from \cite{Nelson2013} (See also Section 2.1 in \citealt{Barraza2021}). For the VSI-unstable disk simulations we run inviscid numerical simulations; therefore, the code solves the Euler equations. However, for the simulations including viscosity, we include the viscous stresses into the hydrodynamical equations, implemented as a parabolic diffusion term in the momentum equation. The viscosity depends on a shear viscosity coefficient ($\nu$), which we set to follow the $\alpha$ viscosity prescription of \citet{Shakura1973}:
\begin{equation}\label{eq:alphashakura}
    \nu=\alpha c_s H,    
\end{equation}
\noindent with $\alpha$ constant through the disks, set to $\alpha = 5\times 10^{-4}$. The quantities $c_s$ and $H$ denote the local sound speed and the disk pressure scale height. For the numerical integration of the diffusion term we chose the Super-Time-Stepping (STS) technique as implemented in \textsc{PLUTO}, which can accelerate the calculations compared to an explicit treatment.

For the simulation grid, the computational domain extends from $0.4$ to $2.5$ code units of length in the radial direction ($r$), and in the azimuthal direction ($\phi$) the grid covers the full $2\pi$ rad. In the meridional direction (colatitude, $\theta$), the grid is set to cover $\sim 10$ disk pressure scale heights at $R=1.0$, that is, $5H$ for each disk hemisphere.
The grid follows a spherical geometry, logarithmically spaced in the radial direction, while evenly spaced in colatitude and azimuth. For the simulations presented, the resolution of the grid is $(r,\theta,\phi)=(512,192,1024)$, which gives a resolution of $\approx 19$ cells per scale height in the meridional direction at $R=1.0$.

We re-scaled the code unit of length of the numerical simulations to $100$ au, which together with the reference aspect ratio at the code unit of length of $H/R=0.1$, sets a model suited for the disk outer regions. Additionally, the stellar mass is set to be equal to $1\, M_{\odot}$. Therefore, we use these reference values to re-scale our simulation results in all figures presented in Section \ref{simulationsresults}, and also for the disk model used in the radiative transfer post-processing (Section \ref{postprocessing}).

In the set of simulations presented in this work, we adopted boundary conditions that consist of enforced zero inflow in $\theta$ and $r$, and an extrapolated density and softened $v_{\phi}$ in the meridional direction (see more details in \citealt{Flock2017} and \citealt{Barraza2021}).
In order to minimize wave reflections close to the inner and outer radial boundaries, we include buffer zones in which the gas density and radial velocity are damped to the initial profiles with a timescale of $10\%$ of the local orbital period. We apply a parabolic damping as introduced by \citet{deValBorro2006}. The radial extents covered by the inner and outer buffer zones are equal to $25\%$ of the grid inner radius and $20\%$ of the outer edge radius, respectively.

\subsection{Planets}

We ran simulation of disks with embedded massive planets for two cases, a Saturn-mass case ($m_p/M_{\odot}=0.3\times 10^{-3}$), and a Jupiter-mass case ($m_p/M_{\odot}=1.0\times 10^{-3}$). For the disk aspect ratio assumed in our simulations ($H/R=0.1$ at the planet's location), the planet masses correspond to $0.3$ and $1.0$ thermal masses, respectively, where the thermal mass is the value at which the planet's Hill sphere matches the sonic point for linear planetary spiral wakes \citep[$\sim 2H/3$, ][]{Goodman2001}, defined as:
\begin{equation}\label{eq:thermalmass}
    M_{th}=\frac{c_s^3}{\Omega_P G} = M_{\star}\left(\frac{H_P}{R_P}\right)^3, 
\end{equation}
\noindent where $\Omega_P$ is the planet's orbital frequency, $M_{\star}$ is the mass of the central star, $H_P$ is the pressure scale height at the planet's position, and $R_p$ is the distance of the planet from the star. The $0.3\,M_{th}$ planet mass case overlaps with the maximum planet mass studied in \cite{Stoll2017} of $100$ Earth masses, while for planet masses equal and above $1\, M_{th}$ non-linear effects are expected to significantly affect the disk structure \citep{Lin1993, Goodman2001}. 

For the inclusion of planets, we use a standard approach of slowly inserting the planet as a gravitational potential of a point of mass, with a smoothing around the location of the planet: 
\begin{equation}\label{eq:planetpotential}
    \Phi_{P}=
            \begin{cases}
              -\frac{GM_{P}}{d},& \textrm{for}\, d\geq d_{rsm},\\
              -\frac{GM_{P}}{d}\left[\left(\frac{d}{d_{rsm}}\right)^4-2\left(\frac{d}{d_{rsm}}\right)^3+2\left(\frac{d}{d_{rsm}}\right)\right],&  \textrm{for}\, d<d_{rsm},\\
            \end{cases}
\end{equation}
\noindent where $d$ is the distance between a fluid element and the planet's position. A potential smoothing length $d_{rsm}$ is used to prevent numerical artifacts at the planet's location. The value of $d_{rsm}$ is set to be three cell diagonals evaluated at the planet location. This corresponds to around $56\%$ of Saturn's and $37\%$ of Jupiter's Hill spheres ($r_{Hill}=r_p(m_p/3M_{\star})^{1/3})$).
In order to avoid numerical artifacts during the inclusion of the planet, its mass is smoothly increased from zero to its final mass in $40$ and $100$ planetary orbits for the Saturn-mass and Jupiter-mass cases, respectively. 

To compare our simulations of VSI-unstable disks directly with a case without VSI, we run the same set of simulations including viscosity following the $\alpha$ prescription of \cite{Shakura1973} (see Eq. \ref{eq:alphashakura}) for a viscosity value comparable to the effective viscosity driven by the VSI \citep{Stoll2017a}. Including viscosity with $\alpha=5\times 10^{-4}$ is enough to damp the VSI in the disk and recover similar structures from previous studies of planet-disk interactions in isothermal disks (e.g., \citealt{Perez2018} and \citealt{Rabago2021}). 

We inspected and post-processed the output after 300 orbits (at $R=1.0$) of evolution for the VSI simulation without a planet, while for the Jupiter and Saturn mass planets simulations we chose the outputs after $145$ and $285$ orbits after the inclusion of the planet, respectively. At these selected times, the planets have already carved a gap and the structures are in a quasi-steady state.\\
A summary of the parameters used in the set of simulations is presented in Table \ref{table:simparams}.

\begin{table}[h!]
\begin{center}
\begin{tabular}{ |c| c|  }
Reference radius & 100 au\\
Aspect ratio at 100 au  & 0.1 \\
Flaring index  & 0.25\\
$\alpha$-viscosity & [$-$; $5 \times 10^{-4}$] \\ 
Stellar Mass [$M_{\odot}$]  & $1.0$ \\
Planet's mass [$m_p/M_{\star}$]  & [$-$; $0.3\times 10^{-3}$; $1.0\times 10^{-3}$] \\  
Density Slope & $-1.0$ \\
Temperature Slope & $-0.5$ \\
Inner Grid Radius  & $0.4$ (40 au) \\
Outer Grid Radius  & $2.5$ (250 au)\\
Grid domain in colatitude & $\pi /2\pm\pi /6.8$ \\
$\#$ of cells in $r$ & 512  \\
$\#$ of cells in ${\theta}$ & 192 \\
$\#$ of cells in $\phi$ & 1024 \\
Inner buffer zone Radius & 0.5 (50 au) \\
Outer buffer zone Radius & 2.1 (210 au) \\
\end{tabular}

\caption{Summary of the parameters used in the global 3D hydrodynamical simulations.}
\label{table:simparams}
\end{center}
\end{table}

\begin{figure*}[htp]
\centering
\includegraphics[angle=0,width=\linewidth]{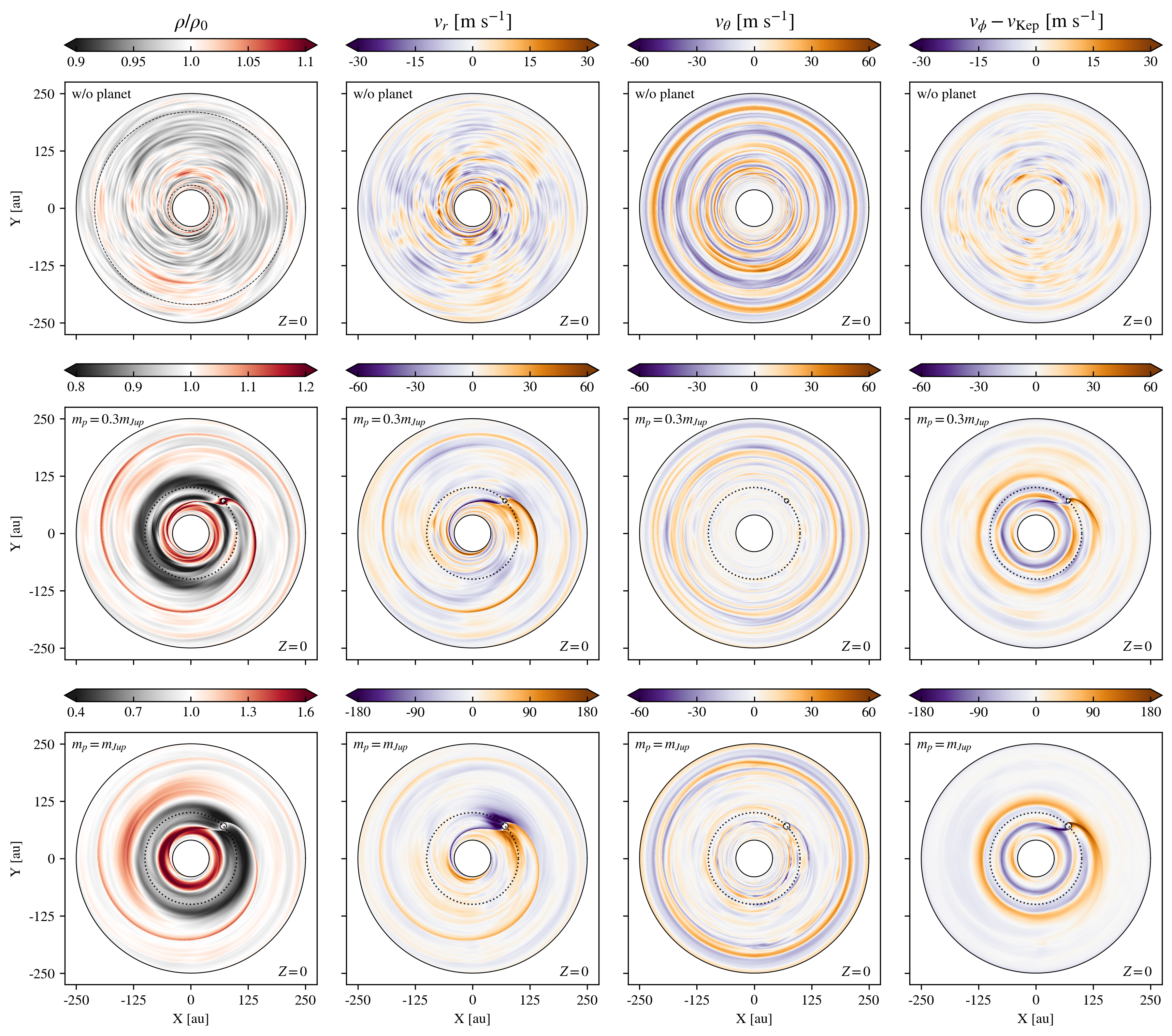}
\caption{Snapshots of the VSI-unstable hydrodynamical simulations at the disk midplane. From top to bottom: simulation without an embedded planet, with an embedded Saturn-mass planet, and with an Jupiter-mass planet. From left to right: gas density relative to the initial value ($\rho/\rho_0$), radial velocity field ($v_r$), meridional velocity field ($v_{\theta}$), and azimuthal velocity deviations from Keplerian rotation ($v_{\phi}-v_{\rm Kep}$). 
The included planets are orbiting at $100$ au from the central star, marked with dotted lines in the second and third rows. The dashed lines in the first panel indicate the buffer zones applied in the hydro simulations.}
\label{fig:simulationsvsimidplane}
\end{figure*}

\begin{figure*}[htp]
    \centering
    \includegraphics[angle=0,width=\linewidth]{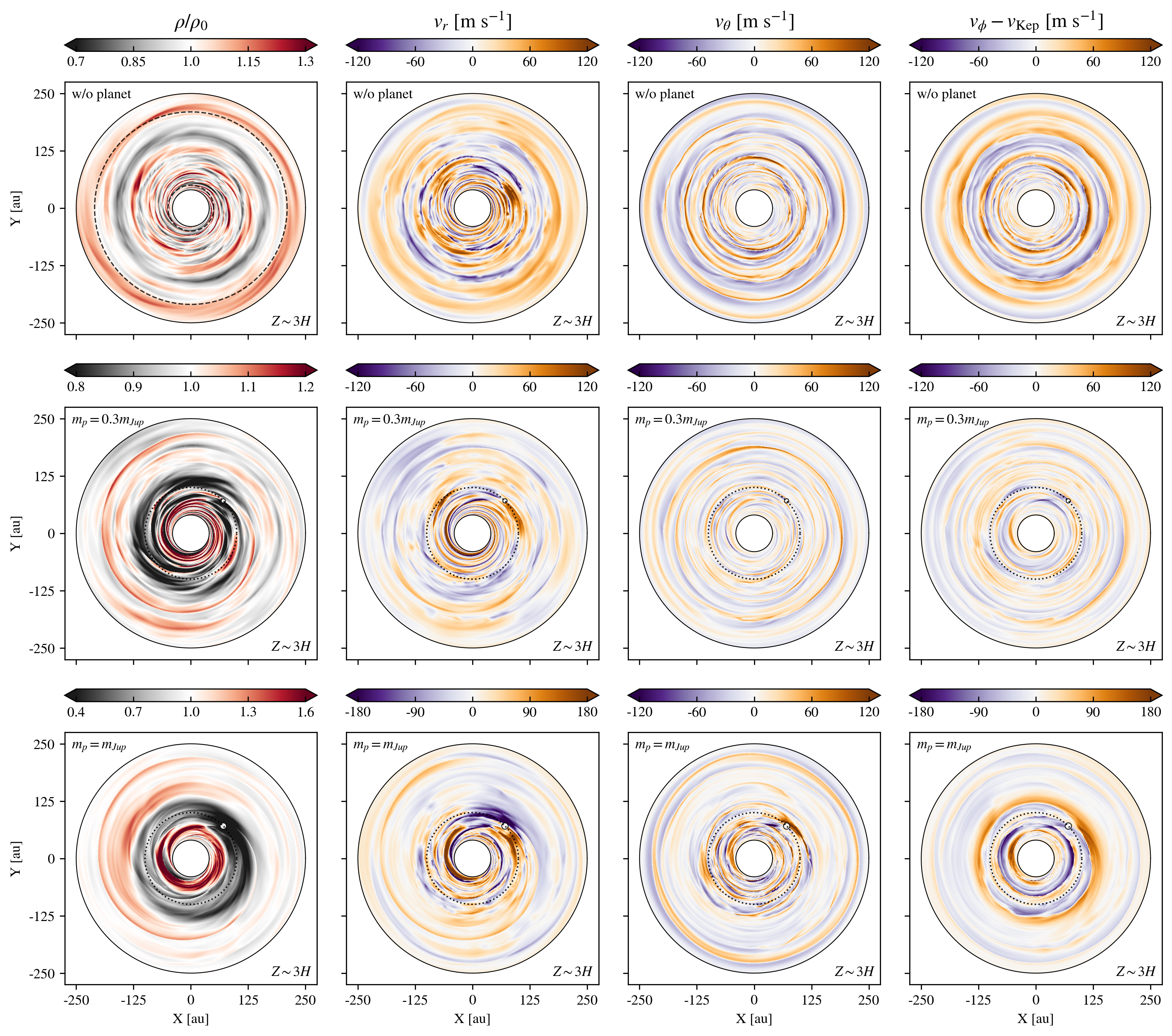}
    \caption{Snapshots of the VSI-unstable hydrodynamical simulations at the disk surface. Same as Figure \ref{fig:simulationsvsimidplane}, but obtained at 3 pressure scale heights from the disk midplane ($Z\sim 3H$).}
\label{fig:simulationsvsisurface}
\end{figure*}

\section{Hydrodynamical Simulations: Results}\label{simulationsresults}

In this section, we present the results of our set of global 3D hydrodynamical simulations.
First, in Section \ref{vsiplanetsimulations} we inspect the simulation outputs of our three VSI-unstable disk simulations, showing the influence of the planets in the VSI-induced velocity and density structures. The comparison of our set of simulations is performed for the disk midplane layer and above 3 pressure scale heights ($Z\sim 3H$).
Particularly important is to inspect different disk heights, since observations of optically thick CO isotopologues, such as $^{12}$CO, trace upper layers of the disk. Optically thinner transitions of less abundant isotopologues, such as C$^{18}$O can trace deeper disk layers (see Figure \ref{fig:tau1surfaces}). In Section \ref{VSIandviscoussimulationresults}, we compare the structures obtained in our VSI-unstable and viscous planet-disk interaction simulations, both in the midplane and surface layers.

\subsection{Simulations of VSI-unstable disks}\label{vsiplanetsimulations}

A comparison of the face-on view of the simulations at the disk midplane is shown in Figure \ref{fig:simulationsvsimidplane}, while Figure \ref{fig:simulationsvsisurface} shows the behavior at three pressure scale heights above the midplane layer. 
In these figures, we present simulations without an embedded planet (first row), with an embedded Saturn-mass planet (second row), and with an embedded Jupiter-mass planet (third row). The columns indicate gas density relative to the initial density field ($\rho/\rho_0$; first column), radial velocity ($v_{r}$; second column), meridional velocity ($v_{\theta}$; third column), and azimuthal velocity deviations from Keplerian rotation ($v_{\phi}-v_{\rm Kep}$; fourth column).
As mentioned above, the simulations have been re-scaled to physical units assuming a central Solar-mass star, and disk radii ranging from 40 to 250 au (1 code unit is 100 au). For reference, the midplane sound speed at 100 au for our assumed setup is $\approx 296$ m s$^{-1}$, $\approx 10\%$ of the local Keplerian speed. For visualization purposes, the colorbar limits are adapted to better cover the perturbation's magnitudes in each panel.
In the gas velocity plots, negative values in the radial direction indicate gas moving towards the central star, in the meridional direction positive values indicate gas flowing downwards (e.g., positive values at $Z\sim 3H$ represent gas moving towards the midplane). In the azimuthal direction positive means the gas is rotating at velocities larger than the Keplerian rotational velocity (i.e., super-Keplerian).
In addition, the location of the planets is indicated by a circle with a radius equal to the planet's Hill radius, and its orbits are indicated by dotted black lines. The dashed black lines in the first panel mark the buffer zones where the parabolic damping is applied.

In the top row of Figures \ref{fig:simulationsvsimidplane} and \ref{fig:simulationsvsisurface}, we recover the characteristic velocity structure of a disk unstable to VSI, dominated by a corrugated circulation pattern \citep[see also, e.g.,][]{Nelson2013,Flock2017,Barraza2021}.
In the disk midplane (Figure \ref{fig:simulationsvsimidplane}), the axisymmetric meridional flows dominate the disk velocity structure, while closer to the disk surface ($Z \approx 3H$; Figure \ref{fig:simulationsvsisurface}) strong velocity perturbations are seen in all three velocity components.
In the simulations including a massive planet (second and third rows in Figures \ref{fig:simulationsvsimidplane} and \ref{fig:simulationsvsisurface}), a gap depleted of gas is carved by the planet, deeper and eccentric for the Jupiter-mass planet case. The density contrast produced by the planet-carved gaps is significantly larger than any of the density perturbations produced by VSI alone, which are also only present in the surface layers.
At the edges and inside the planet-carved gaps, rings of super- and sub-Keplerian gas are seen, while similar non-Keplerian flows are induced by the VSI-induced density perturbations at the surface layers; although, with a corrugated morphology.
Additionally, asymmetric structures are triggered by the planet: spiral arms via Lindblad resonances are induced by the planets in the density, radial velocity and azimuthal velocity fields, clearly seen at the disk midplane (see also Figure \ref{fig:alphavsvsimidplane}). Around the planet's location, strong planetary spiral wakes are also produced by the massive planets. Finally, a large-scale vortex is produced at the outer edge of the gap carved by the Jupiter-mass planet, seen as a horseshoe-shaped gas overdensity; however, less prominent in the disk velocities. While VSI also induces asymmetries, as multiple vortices are also present in the disk without planets, these have smaller size scales compared to the Jupiter-induced one for the examined outputs.
At the disk midplane, we observe that the planet-induced perturbations dominate the overall disk structure for $\rho$, $v_r$ and $v_{\phi}$. Interestingly, in the meridional velocities, corrugated meridional flows induced by the VSI unstable modes are still significant, however, only in the outermost region of the disk. Such velocity structure is a consequence of the damping of the VSI produced by the presence of the massive planets being more efficient at the midplane layer. The efficient damping towards the disk midplane can also be linked to the vertical shear rate ($R \partial\Omega/\partial z$) increasing with disk height, therefore, maintaining stronger VSI motions towards the disk upper layers. The planet-induced damping of the VSI is apparent in the region inside the planet's radial location, along the planetary gap, and also at the gap's outer edge. 
In the disk's upper layers, damping of the VSI meridional flows seems still to be present for the Jupiter-mass case, whereas for the Saturn-mass case a mixture of VSI-induced structures and the planet-induced spiral arms is observed. 
The structure of $v_{\theta}$ and $v_{\phi}$ at $Z\approx 3H$ for the simulations without a planet and a Saturn-mass planet appears to be similar overall. Global damping of the VSI-induced flows is produced by the Saturn planet, inducing perturbations that reach lower velocities for all components.
Finally, in the disk surface layers influenced by the Jupiter planet, the localized velocity flows around the planet's location are the strongest features in the radial and meridional directions, while in the azimuthal direction a ring of Super-Keplerian gas at the outer gap edge is the most prominent velocity perturbation.
From kinematic observations, identifying the symmetric and asymmetric velocity and density structures in a resolved view of the disk is required to separate different scenarios (see Section \ref{rtresults}).\\

\begin{figure*}[htp!]
    \centering
    \includegraphics[width=\textwidth]{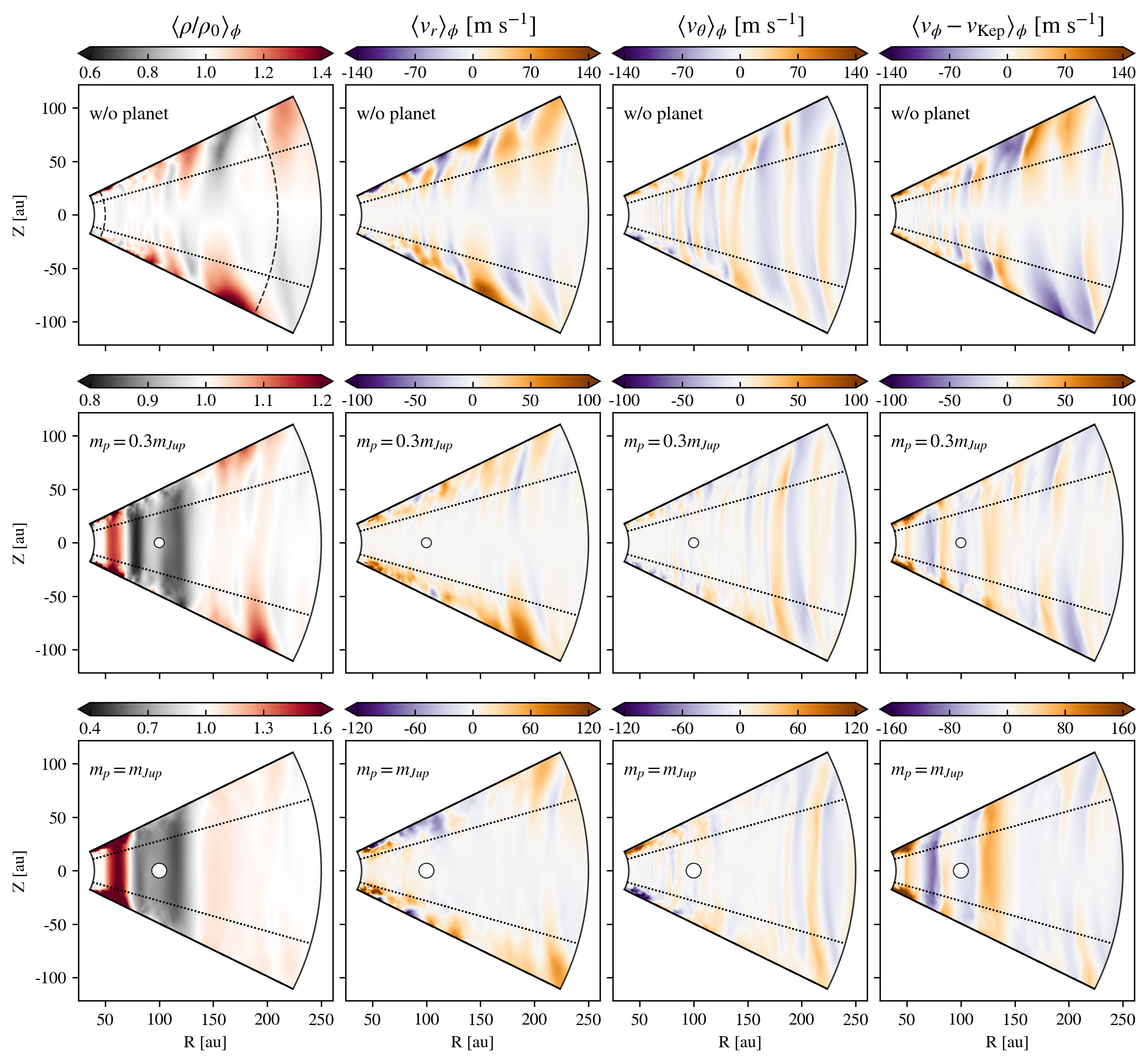}
    \caption{$R-Z$ view of the azimuthally-averaged fields of our VSI-unstable disks simulations. From left to right, gas density normalized by the initial value, radial velocity ($v_r$), meridional velocity ($v_{\theta}$), and azimuthal velocity deviations from Keplerian rotation ($v_{\phi}-v_{\rm Kep}$). From top to bottom, simulation without an embedded planet, with a Saturn-mass planet ($m_p=0.3m_{Jup}$), and a Jupiter-mass planet ($m_p=m_{Jup}$). The black dashed lines in the first panel show the simulation buffer zones. The black dotted lines indicate the height in which $Z\approx 3H$. The white circles show the radial location of the planets, where each circle's diameter is set by the planet's Hill sphere.
    }
    \label{fig:azimuthalaverages}
\end{figure*}

To highlight the stronger damping of the VSI at the disk midplane produced by the presence of the embedded planets, we show a $Z-R$ view of our set of VSI-unstable disk simulations in Figure \ref{fig:azimuthalaverages}. We present the azimuthally-averaged fields, following the same order of presentation as the panels of Figures \ref{fig:simulationsvsimidplane} and \ref{fig:simulationsvsisurface}.
The vertical sliced view of the disk gas velocities shows that the damping of the VSI is more effective in the region below three pressure scale heights from the disk midplane, marked by black dotted lines. In the outermost regions of the disk flows induced by the VSI are still active, characterized by columns of gas moving upwards or downwards. A sketch of the meridional velocity structure in a VSI-unstable disk with an embedded massive planet is presented in Figure \ref{fig:vsiplanetsketch}.
The different symmetry of the flow direction with respect to the midplane can be exploited to separate between VSI- and planet-induced perturbations (discussed in Section \ref{vsiplanetsdiscussion}).
In the radial and azimuthal directions, the VSI flows are symmetric with respect to the disk midplane, while planet-induced flows are anti-symmetric. On the contrary, in the meridional direction the VSI flows are anti-symmetric with respect to the midplane, while planet-induced flows are symmetric; that is, at a particular radius, gas moves towards the midplane or away from the midplane at both disk hemispheres.
We isolate such an effect in Figure \ref{fig:simradialprofiles}, showing the radial profiles of the velocity at $Z\sim 3H$ from the midplane for both disk hemispheres. A clear difference in symmetry relative to the midplane is seen from the (anti-)correlations of the flow directions.
Lastly, we stress that, while a net meridional flow towards the gap carved by the massive planet is seen in the $v_{\theta}$ azimuthal averages (Figures \ref{fig:azimuthalaverages} and \ref{fig:simradialprofiles}), they are relatively weak even at $Z\sim 3H$. Such low magnitudes are consistent with larger planet masses needed to explain observed meridional flows towards gaps \citep[$m_p \geq 2 M_{Jup}$, e.g.,][]{Teague2019c}. Moreover, these planet-induced meridional flows are not of an axisymmetric morphology in the $r-\phi$ plane for VSI-turbulent disks. Such characteristic morphology might be of importance when interpreting resolved 2D maps of the line-of-sight velocity (e.g., Figures \ref{fig:residualsraw12co} and \ref{fig:residualssimobserve}, see following Section \ref{VSIandviscoussimulationresults}).

\begin{figure*}[htp!]
\centering
\includegraphics[angle=0,width=\linewidth]{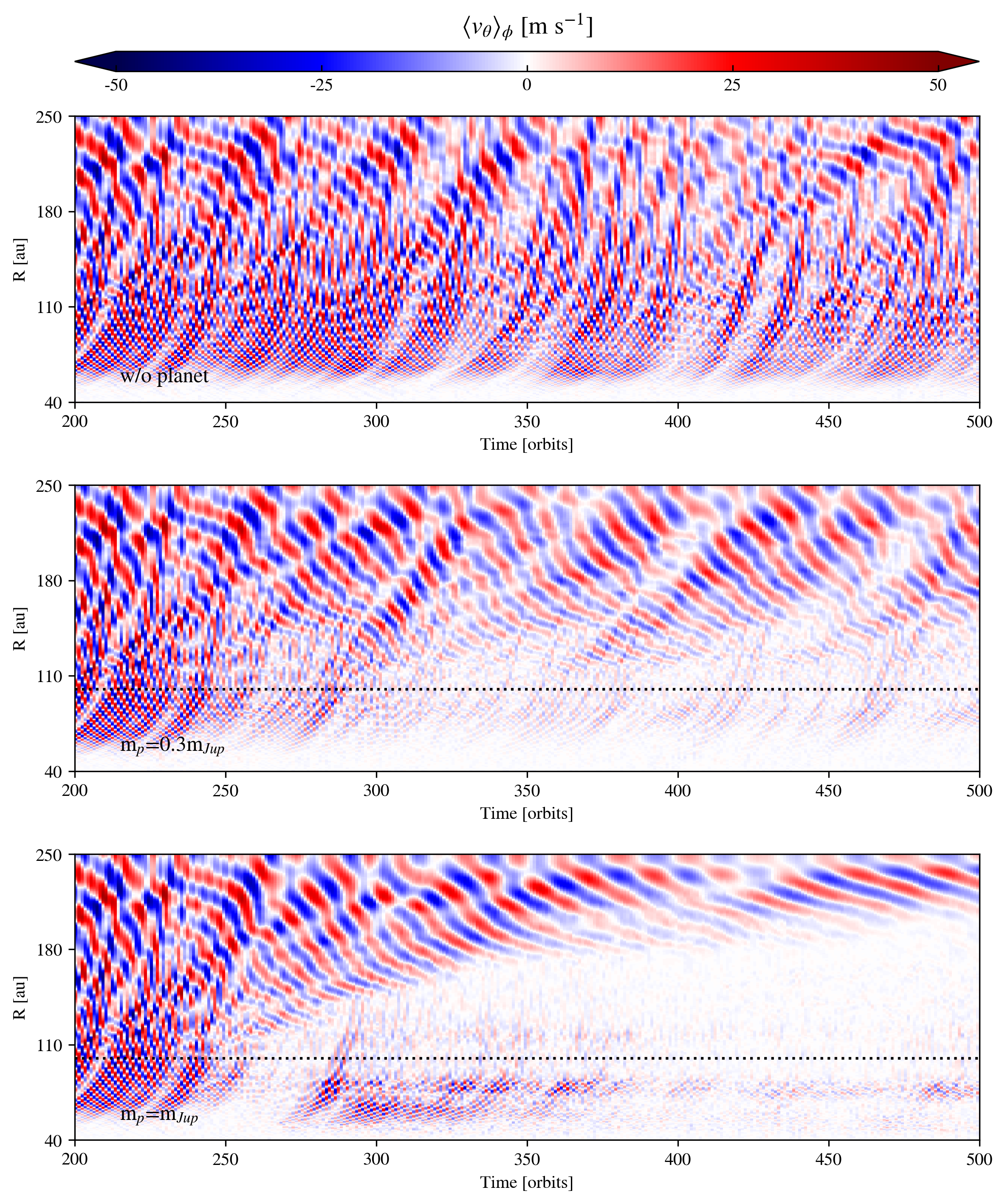}
\caption{Time evolution of the azimuthally averaged meridional velocity at the disk midplane ($Z=0$). From top to bottom, we show simulations of VSI-turbulent disks without an embedded planet, with an embedded Saturn-mass planet, and with an embedded Jupiter-mass planet. The orbits are measured at the planets' radial location ($100$ au), shown by black dotted lines. The figure x-axis starts at the time at which the planets are inserted in the simulations.}
\label{fig:midplanetimeevol}
\end{figure*}

To visualize the planet-induced damping of the VSI, we explored the time evolution of meridional velocity perturbations in our VSI-unstable disk simulations. In Figure \ref{fig:midplanetimeevol}, we show the azimuthal average of the midplane meridional velocity ($\langle v_{\theta} \rangle_{\phi}$ at $Z=0$) at each orbit, for $300$ planetary orbits starting at the time when the planets are included in our planet-disk interaction simulations. The axisymmetry of the VSI unstable modes in the azimuthal direction allows us to follow the mode evolution in the azimuthal averages.
In the first row of Figure \ref{fig:midplanetimeevol}, we show the time evolution of a simulation without an embedded planet, in which the VSI is operating in its saturated state. Radial migration of the VSI modes towards the central star is observed, on top of narrow radial regions of low velocities that migrate outwards. These results are consistent with previous findings on the time evolution of VSI unstable modes \citep[e.g.,][]{Stoll2014, Pfeil2021, Svanberg2022}. Note that the velocities close to the inner grid edge are damped by the effect of the simulation buffer zones.
In the second and third rows of Figure \ref{fig:midplanetimeevol}, we show the time evolution of the VSI unstable simulations with an embedded Saturn-mass planet and the simulation with an embedded Jupiter-mass planet, respectively. The planets are in orbit at 100 au from the central star, indicated by the horizontal black dotted line. From the weakened meridional velocity perturbations, we observe that the planets produce a damping of the meridional flows induced by the VSI, particularly strong in the regions inside its radial orbit, along the gap region and gap outer edge. The damping produced by the Saturn-mass planet is less efficient than for the Jupiter case. The VSI motions are still vigorous in most of the outer regions of the disk after 300 orbits ($r\gtrsim 150$ au), and show an apparent convergence to a steady state. Due to its stronger influence on the disk structure, the embedded Jupiter produces a more effective damping of the VSI, in which the VSI motions are damped completely up to $\sim 200$ au from the star. Contrary to the Saturn-mass simulation, the Jupiter case has yet to fully converge by the end of our simulation ($300$ planetary orbits), where the damped region could still grow in radius. A longer simulation run with a larger radial domain is needed to further study the steady state of the gas dynamics of VSI-unstable disk with a Jupiter planet.\\
In order to confirm that the smaller values of the azimuthally-averaged meridional velocity are not exaggerated due to a break of the VSI axisymmetry by the planet-disk interactions, we validated this result by exploring the time evolution of the azimuthally-averaged absolute value of $v_{\theta}$, showing the same behavior as presented above.\\
Our results of planet-induced VSI damping are consistent with previous findings by \cite{Stoll2017} and independent simulations by \cite{Ziampras2023} and \cite{Hammer2023}, in which we found a stronger damping of VSI motions for a more massive planet. These results can be extrapolated to planets with larger masses, in which planets above one thermal mass (equal to one Jupiter mass for our disk model, see Eq. \ref{eq:thermalmass}) would strongly damp the VSI and dominate the overall disk gas dynamics. Moreover, our results are consistent with the findings of \cite{Lehmann2022}, where VSI is weakened inside disk pressure bumps.\\
Regarding the origin of the damping, we did not find a direct correlation of the dampened regions with other quantities. However, the time scale of the damping matches the times-scale of the gap opening by the planets. Previously, \cite{Stoll2017} attributed the damping to the formation of vortices. On the contrary, \cite{Manger2018} found that VSI triggers and coexists in large-scale anti-cyclonic vortices \citep[see also, ][]{Hammer2023}. While we did not explore this further, the influence of the planet on the vorticity field might play an important role in the VSI damping, by creating ring structures on the vorticity field. The origin of damping is hard to isolate, since the planets have significant effects on the gas density and pressure structure, from the gap opening and the launching of Lindblad spirals. Therefore, we can only conclude that the VSI is affected by a combination of the effects mentioned above.\\
Finally, the planet-induced damping of the VSI can substantially impact the settling of dust grains towards the midplane, where the damping is strongest. Therefore, a global lower dust scale height is expected for VSI-turbulent disks with embedded planets, relative to a disk with VSI alone, which would vertically mix solid particles \citep[e.g.,][]{Flock2020}. However, for particular dust rings at the outer edges of planetary gaps the vertical mixing would depend on the planet's mass, since planets massive enough to create meridional flows can also lift up dust pebbles \citep{Bi2021, Binkert2021}.
In addition, the VSI damping can strongly modify the turbulent stresses produced by VSI, which define its ability to transport angular momentum \citep[see][]{Ziampras2023}. For kinematic observations of rotational lines, it is expected that the planet-induced damping reduces the chances of detecting VSI signatures near a planetary-gap region, especially relevant for molecules tracing layers near the disk midplane (e.g., C$^{18}$O(3-2), see Section \ref{rtresults}).

\subsection{VSI-unstable disks vs $\alpha$-viscous disks}\label{VSIandviscoussimulationresults} 

\begin{figure*}[htp!]
    \centering
    \includegraphics[width=\textwidth]{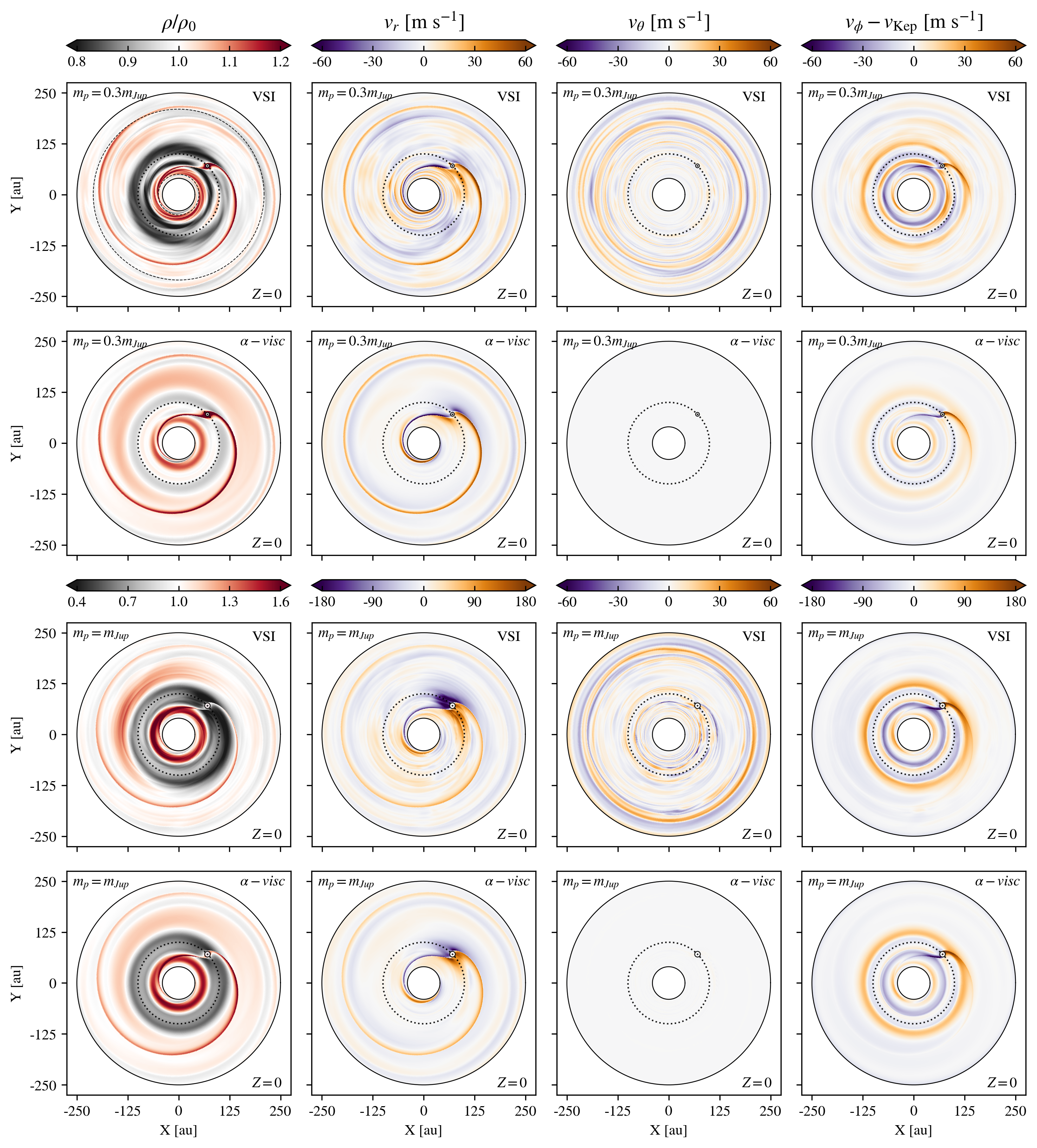}
    \caption{Outputs of VSI-unstable disk and $\alpha$-viscous disk ($\alpha=5\times 10^{-4}$) simulations with embedded massive planets at the disk midplane ($Z=0$). From left to right, gas density normalized by the initial value, radial velocity ($v_{r}$), meridional velocity ($v_{\theta}$), and azimuthal velocity deviations from Keplerian rotation ($v_{\phi}-v_{\rm Kep}$). From top to bottom, VSI simulation with a Saturn planet ($m_p=0.3m_{Jup}$), $\alpha$-viscous simulation with a Saturn planet, VSI simulation with a Jupiter planet ($m_p=m_{Jup}$), and $\alpha$-viscous simulation with a Jupiter planet. The black dashed lines in the first panel show the simulation buffer zones. The black dotted lines show the planets' orbital radius. The white circles show the location of the planets. The circles' diameters are equal to the planets' Hill spheres.
    }
    \label{fig:alphavsvsimidplane}
\end{figure*}

\begin{figure*}[htp!]
    \centering
    \includegraphics[width=\textwidth]{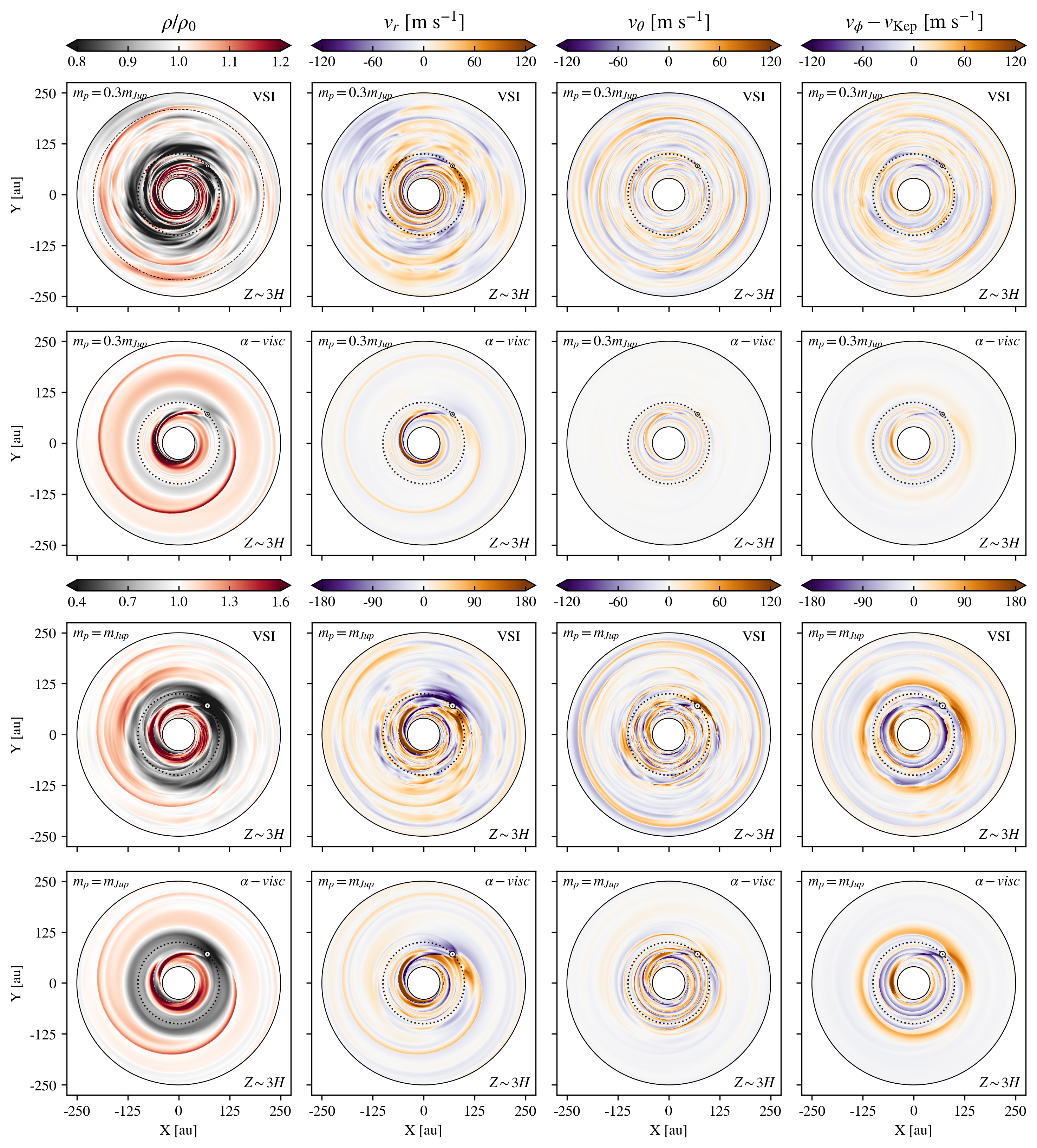}
    \caption{Outputs of VSI-unstable and $\alpha$-viscous disk ($\alpha=5\times 10^{-4}$) simulations with embedded massive planets at the disk surface ($Z\sim 3H$). Same as Figure \ref{fig:alphavsvsimidplane}, but for a disk slice three pressure scale heights above the disk midplane ($Z\sim 3H)$.
    }
    \label{fig:alphavsvsisurface}
\end{figure*}

Intending to highlight the structures resulting from the interplay of VSI and massive planets, we compare turbulent VSI-unstable disk simulations against viscous $\alpha$  disks simulations. As above, we run the cases of embedded Saturn-mass and Jupiter-mass planets, examining the perturbed density and velocity fields for both the midplane layers (Figure \ref{fig:alphavsvsimidplane}), and surface layers ($Z\approx 3H$; Figure \ref{fig:alphavsvsisurface}). Comparing these sets of simulations is a simplified approach to contrast planet-disk interactions with and without VSI operating in the disk, in which the viscosity included in the $\alpha$ models prevents the growth of the VSI \citep[see also, ][]{Stoll2017}.
We present the face-on view of the gas density and velocities in the same order of columns as Figures \ref{fig:simulationsvsimidplane}, \ref{fig:simulationsvsisurface} and \ref{fig:azimuthalaverages}. While the first and third rows display the VSI-unstable disks, overlapping with the results shown in Figures \ref{fig:simulationsvsimidplane} and \ref{fig:simulationsvsisurface}, the second and fourth rows display the $\alpha$ disk simulations outputs for $\alpha = 5 \times 10^{-4}$ (see Section \ref{hydrosimulations}).\\
From the direct comparison, it is clear that the VSI induces additional fine structure in all velocity fields, while the structures in the $\alpha$-disk simulations are smoothed by the viscous diffusion. The simulations including $\alpha$ viscosity show slightly lower velocity magnitudes of the flows localized around the planet, and the super-Keplerian ring at the outer edge of the gap induced by the planet. From the presence of a less depleted gap in the $\alpha-$disk models, better seen in the Saturn-mass case, it is certain that these differences are the result of the VSI effective turbulent $\alpha$ being slightly smaller than the value set for the $\alpha$-disk simulations. Due to the difficulty of entirely suppressing VSI motions in an $\alpha-$disk with a lower $\alpha$ value in locally isothermal simulations, we concentrate on the morphological differences of the coherent large-scale motions, with potential distinct observational signatures in kinematic CO line observations (see Section \ref{rtresults}).\\
Differences between VSI and $\alpha$-disk simulations are seen in the meridional velocity structure, in which the additional meridional quasi-axisymmetric rings induced by the VSI are present, evident in the midplane and surface of the disk. Moreover, inside the gap carved by the Jupiter planet the VSI-induced turbulence disrupts the meridional flow structure at the surface layers, contrary to the smoother ringed flows along the gap in the viscous case.
In the radial direction, the VSI adds additional spiral-like perturbations at $Z\approx 3H$, likely from the interaction between VSI-unstable modes and the Lindblad spirals driven by the planets. Here the VSI also disrupts the Lindblad spiral triggered by the Saturn-mass planet, creating arc-like features.
Finally, in the azimuthal velocities, the VSI induces additional sub- and super-Keplerian rings at the outer disk surface layers, evident for the Saturn-mass case. In addition, we observe that in the VSI-unstable case the Jupiter planet triggers a strong anti-cyclonic vortex at the outer edge of the gap, also visible in the perturbed gas density. Such difference might be caused by the fact that the effective VSI turbulent $\alpha$ is slightly smaller than the assumed value of $5\times 10^{-4}$ for the viscous disk. Additional simulations assuming lower constant $\alpha$ could solve this discrepancy. Such a simulation would require an alternative method to damp the VSI.\\

\section{Radiative Transfer and Simulated Observations: Methods}\label{postprocessing}

\subsection{Radiative Transfer Setup}\label{rtsetup}

To produce synthetic images of molecular line emission of our set of hydrodynamical simulations, the outputs are post-processed with the Monte-Carlo radiative transfer code \textsc{radmc-3d\footnote{\url{http://www.ita.uni-heidelberg.de/~dullemond/software/radmc-3d}}} \citep{Dullemond2012} version 2.0.
We constructed the \textsc{radmc-3d} input files from the simulation data following the procedure described in \cite{Barraza2021}. The scripts to construct the \textsc{radmc-3d} input files were partially based on an early version of \textsc{fargo2radmc3d\footnote{\url{https://github.com/charango/fargo2radmc3d}}} \citep{Baruteau2019}, and \textsc{radmc3dPy\footnote{\url{https://www.ita.uni-heidelberg.de/~dullemond/software/radmc-3d/manual_rmcpy/}}}. The observables explored in this paper are the spatially resolved velocity centroid maps (also labeled as line-of-sight velocity maps), computed from synthetic CO line emission data cubes with good velocity resolution (see Section \ref{kinematictools}). 
We compute the synthetic data cubes for three different CO isotopologues: $^{12}$CO, $^{13}$CO and C$^{18}$O, in order to study the effect of probing different disk layers (see Figure \ref{fig:tau1surfaces}).
Particularly, we compute predictions for the J=3-2 rotational transition observable within ALMA Band 7. Our selection is motivated by the better spectral resolution available in Band 7 than for the J=2-1 transition (within Band 6); therefore, this is better suited for characterizing the velocity structure in kinematic observations. Nonetheless, predictions for the J=2-1 transition would result in an equivalent outcome \citep[see][]{Barraza2021}.\\
As mentioned above, we used a similar model setup as presented in \cite{Barraza2021} to use the outputs of the simulations as inputs into the radiative transfer code. We re-scaled the simulation output radial grid to $R_0=100$ au (i.e., one code unit is re-scaled to $100$ au), and assumed a $1 M_{\odot}$ central star. We volume-averaged the simulation data onto a coarser grid, halving the grid resolution in each direction to speed up the radiative transfer calculations. We also extended our disk, including an inner disk that follows the equilibrium solution used as the initial condition in the simulation, which goes from 10 au to the simulation grid's inner edge of 40 au. Therefore, our full disk radiative transfer model extends from 10 to 250 au. Additionally, we removed the cells adjacent to the grid edges in colatitude, to prevent tracing the grid cells affected by boundary conditions. However, for the assumed gas density of the model, in which the total gas mass of the disk in molecular hydrogen is $0.05\,M_{\odot}$, the layers traced by the explored CO isotopologues are unlikely to be affected by the dynamics close to the boundaries in $\theta$, as shown in Figure \ref{fig:tau1surfaces}.\\
The disk temperature is computed via dust thermal Monte Carlo radiative transfer. It is assumed that gas and dust have the same temperature. Since our hydrodynamical simulations only treat the gas dynamics, the dust is included manually adopting a gas-to-dust mass ratio of $100$ through the disk.
The dust is composed of a mixture of astrosilicates, amorphous carbon, and vacuum. The optical constant of the mixture was calculated using \textsc{optool\footnote{\url{https://github.com/cdominik/optool}}} \citep{Dominik2021}, applying the Bruggeman mixing formula and Mie theory to compute the dust opacities \citep{Bohren1983} (see, e.g., \citealt{Baruteau2019} and \citealt{Barraza2021}). The computed opacities have a resulting dust intrinsic density of $2.0\,\rm g\,cm^{-3}$ \citep[see further details in ][]{Brown2021}. We adopted a highly simplified dust structure in order to speed up the calculations by assuming only one representative dust size bin for grain sizes between $0.01\,\mu\textrm{m}$ and 10 $\mu\textrm{m}$, following the same distribution as the gas.
Our results are not significantly affected by these assumptions, since small grains dominate the resulting temperature structure, and we do not include dust in the image ray-tracing.\\
For the calculations, we assume that the central star radiates as a perfect black body with an effective temperature of $T_{\star}=7000\,\textrm{K}$ and a radius of $R_{\star}=1\,R_{\odot}$. We use $10^9$ photon packages to compute the dust temperature via thermal Monte Carlo radiative transfer including absorption and scattering opacities (assuming Henyey-Greenstein anisotropic scattering), while $10^8$ photon packages are used for the image ray-tracing. For all the presented images, we assume a distance to the source of 100 pc.\\
For the molecular abundances, a constant fraction of $^{12}$CO relative to H$_2$ of $1\times10^{-4}$ its assumed for the entire disk, while for $^{13}$CO and C$^{18}$O the $^{12}$CO abundance its scaled by $\sim 77^{-1}$ and $\sim 560^{-1}$ (see Section 3.1 in \citealt{Barraza2021}). The line emission is computed assuming LTE, using molecular data from the \textsc{LAMDA\footnote{\url{https://home.strw.leidenuniv.nl/~moldata/}}} database \citep{Schoier2005}. Variations of CO abundance from photo-dissociation are not included in our models, while a simplified CO freeze-out is included by reducing the CO abundance by a factor $10^{-5}$ in cold regions ($T\leq 21$ K).
The synthetic data cubes are computed with a fine velocity resolution of 10 m s$^{-1}$. Then, the channels are averaged to obtain data cubes with a coarser resolution, matching a velocity resolution of $100$ m s$^{-1}$, observable with ALMA. This procedure mimics telescope limitations without the need to include artificial micro-turbulence in the radiative transfer model. Images of an individual RAW (without spatial convolution) $^{12}$CO channel map for the different models are presented in Figure \ref{fig:channelmaps}, while a view of all channels is included as supplementary material. The synthetic data cubes are used to simulate ALMA observations.\\
A summary of the parameters used in the radiative transfer predictions is compiled in Table \ref{tab:rtparams}.

\begin{table}[h!]
\begin{center}
\begin{tabular}{ |c| c|  }
Stellar radius & $ 1.0\,R_{\odot}$\\
Stellar effective temperature & $7000\,\textrm{K}$\\
Distance & 100 pc\\
Disk total gas mass & 0.05 $M_{\star}$\\
Disk total dust mass  & $5\times10^{-5}$\\
Maximum dust size & 10 $\mu$m\\
Minimum dust size & 0.01 $\mu$m\\
Dust size slope & -3.5  \\
Dust intrinsic density & 2.0 g cm$^{-3}$ \\
Disk inclinations & [$5^{\circ}$,$15^{\circ}$,$30^{\circ}$]\\
\end{tabular} 
\caption{Summary of the parameters used in the radiative transfer predictions.} 
\label{tab:rtparams}
\end{center}
\end{table}

\subsection{Simulated Observations}\label{methodssimobserve}

To predict how the synthetic images of our models would appear in an interferometric ALMA observation, we simulated mock observations with the Common Astronomy Software Applications package (\textsc{CASA}\footnote{\url{https://casa.nrao.edu/index.shtml}} version 6.4; \citealt{McMullin2007}). 
The simulated observations are computed in three steps: simulation of the observed visibilities, inclusion of noise, and cleaning of the dirty image. First, we use the task \texttt{simobserve} to simulate the observed visibilities using our RAW synthetic data cubes as input images. The uv-coverage is computed for a combination of two antenna configurations, one extended (C-7, with longest baselines of 3.6 km) and one compact (C-4, with longest baselines of 784 m). The adopted antenna configuration is similar to the configuration of the large program MAPS \citep[e.g.,][]{Oberg2021}; however, for ALMA Band 7, resulting in a spatial resolution of $84\times62$ mas ($8.4\times 6.2$ au). The simulated visibilities consider an integration time on the compact configuration of $25\%$ of the extended one (10h and 2.5h, respectively). The combination of two different antenna configurations cover both short and long baselines, in order to recover information from large and small spatial scales, respectively. Relatively long on-source integration times are used in the set of simulated observations, necessary to have good uv-coverage, crucial for a final image with the fidelity to extract the kinematic information. Furthermore, such long integration is also needed due to the difficulty of getting fairly good signal-to-noise in high-resolution (spatial and spectral) CO observations. 
Second, we use the task \textsc{sm.corrupt()} (simulator.corrupt) to corrupt the simulated data, adding errors in the visibilities. We include errors with an RMS of 1 mJy/beam per channel corresponding to our assumed long integration times, calculated with the ALMA sensitivity calculator\footnote{\url{https://almascience.eso.org/proposing/sensitivity-calculator}}. These noise calculations assume particular atmospheric conditions, that significantly affect the time required to reach a particular sensitivity.
Third, we applied \textsc{CASA} \texttt{tclean} to reconstruct the image from the modeled dirty image visibilities, following the CLEAN algorithm. In this process, we used the multi-scale mode and a briggs weighting scheme. Moreover, the cleaning was performed using non-interactive Automasking (auto-multithresh; \citealt{Kepley2020}), which automatically generates the masks used during the process. Such automatic masking is possible due to the known morphology of the emission from the radiative transfer models; however, in real observations masking the image manually is still recommended. As a final product, a cleaned spectral cube with the expected artifacts from a real ALMA observation is obtained. Further details on our method to simulate ALMA observations are presented in Section \ref{simulatedkinematicsignatures}.

\subsection{Kinematic Analysis Tools}\label{kinematictools}

The kinematic signatures of the simulated observations are extracted in two steps. First, the line-of-sight velocity map is computed from the data cube. Second, the best fit Keplerian model to the line-of-sight velocity map is determined. This is then subtracted from the original velocity map to reveal coherent non-Keplerian gas flows.
For the first step, we compute velocity centroid maps (v$_0$) from the data cubes using a Gaussian function to fit the CO line emission in each pixel of the collapsed image. For this we use the publicly available Python package \textsc{Bettermoments\footnote{\url{https://github.com/richteague/bettermoments}}} \citep{Teague2018c, Teague2019b}. This package robustly computes the centroid maps of spectral line data for a variety of methods, and their respective statistical uncertainties. A Gaussian function is chosen as it gives the best results for our particular set of synthetic models. We extract the disk velocity perturbations from the map of velocities projected into the line-of-sight. For this purpose we use the Extracting Disk DYnamics Python suite \textsc{Eddy\footnote{\url{https://github.com/richteague/eddy}}} \citep{eddy} to obtain the best fitting Keplerian disk model for the $^{12}$CO(3-2) velocity centroid maps of the simulated ALMA observations. 

We used a model that assumes a geometrically thick disk with an elevated emission surface, in which the emitting surface is parameterized by:
\begin{equation}
   z(r) = z_0 \times \left(\frac{R}{1^{\prime\prime}}\right)^{\psi} \times \exp\left(-\left[\frac{R}{R_{\rm taper}}\right]^{q_{\rm taper}}\right),
\end{equation}
\noindent where $R$ is the disk cylindrical radius in arcseconds, $\psi$ dictates the flaring of the emission surface, and $z_0$ and $R_{\rm taper}$ the reference disk aspect ratio at 1 arcsecond and exponential taper reference radius in arcseconds. However, for our fitting we assumed the limit $R_{\rm taper}=\infty$, which better fits our disk models, also reducing the number of free parameters.

For the fitting of the disk rotation, we assume that the rotation curve follows a Keplerian profile accounting for the altitude of the emission height:
\begin{equation}
    \textrm{v$_{\rm Kep}$} = \sqrt{\frac{GM_{\rm star}R^2}{(R^2+z^2)^{3/2}}},
\end{equation}
\noindent with $M_{\rm star}$ the mass of the central star. In this case, the cylindrical radius $R$ and the emission surface height $z$ are in meters, adapted using the distance to the source in our model of 100 pc. In the following, the disk velocity model is projected into the line of sight considering the contribution of the azimuthal velocity component only:
\begin{equation}
    \textrm{v$_{mod}$} = \textrm{v$_{\rm Kep}$}\cdot \cos{\phi} \cdot \sin{i}+\textrm{v$_{\rm LSR}$},
\end{equation}
\noindent where $\phi$ is the polar angle of the image pixel (measured east of north relative to the red-shifted major axis) and v$_{\rm LSR}$ is the systemic velocity, set to zero in our models. 

For the fitting procedure, we fix the disk inclination to the input model inclination and the distance to the system to $100$ pc, and considering as free parameters $M_{\rm star}$, disk PA, $z_0$, $\psi$,v$_{\rm LSR}$, $x_0$ and $y_0$. Then, a series of MCMC chains are run to find the best-fit parameters of the geometrically thick Keplerian disk model. For this paper, we used 128 walkers that take 2000 burn-in steps and additional 500 steps to sample the posterior distributions for the model parameters. A delimited radial region of the disk is considered in the model fitting, set to [0.55,2.0], [0.58,1.85] and [0.6,1.7] arcseconds for inclinations of 5, 15 and 30 degrees, respectively.
Finally, the velocity perturbations are extracted by subtracting the velocity centroid map of the best-fit disk model (v$_{mod}$) from the original (v$_0$).

An alternative way to look at the disk kinematic structure is to obtain an azimuthally averaged view of the disk velocities (radial profiles); however, in this paper we only studied the two-dimensional view of the deviations from Keplerian rotation. In principle, the axis-symmetry of the VSI flows could be exploited with such approach, while also boosting the signal-to-noise of the simulated observations, reaching higher precision in velocity. Unfortunately, a degeneracy between the flows produced by the VSI and a massive planet might be faced when exploring the radial velocity profiles of the upper $^{12}$CO(3-2) emission layer only, as suggested by our simulations (see Figure \ref{fig:simradialprofiles}). Moreover, the extraction of the radial profiles is extremely sensitive to systematic errors and more computationally expensive. Nevertheless, looking at the velocity radial profiles has enormous potential to unravel VSI motions, by allowing further exploration of flow correlations among velocity components for both disk layers (see Section \ref{vsiplanetsdiscussion}).\\
Finally, we highlight that alternative tools to \textsc{Bettermoments} and \textsc{Eddy} are also available, such as GMoments\footnote{ \url{https://github.com/simoncasassus/GMoments}} \citep{Casassus2019} and Discminer \citep{Izquierdo2021a, Izquierdo2021b}. Differences between methods can be found in terms of the flexibility of the models, specific features, and varying performance for particular targets (see, e.g., \citealt{DiskDynamics2020}).

\section{Radiative Transfer and Simulated Observations: Results}\label{rtresults}

\subsection{Kinematic signatures: An idealistic view}

\begin{figure*}[htp!]
    \centering
    \includegraphics[width=\textwidth]{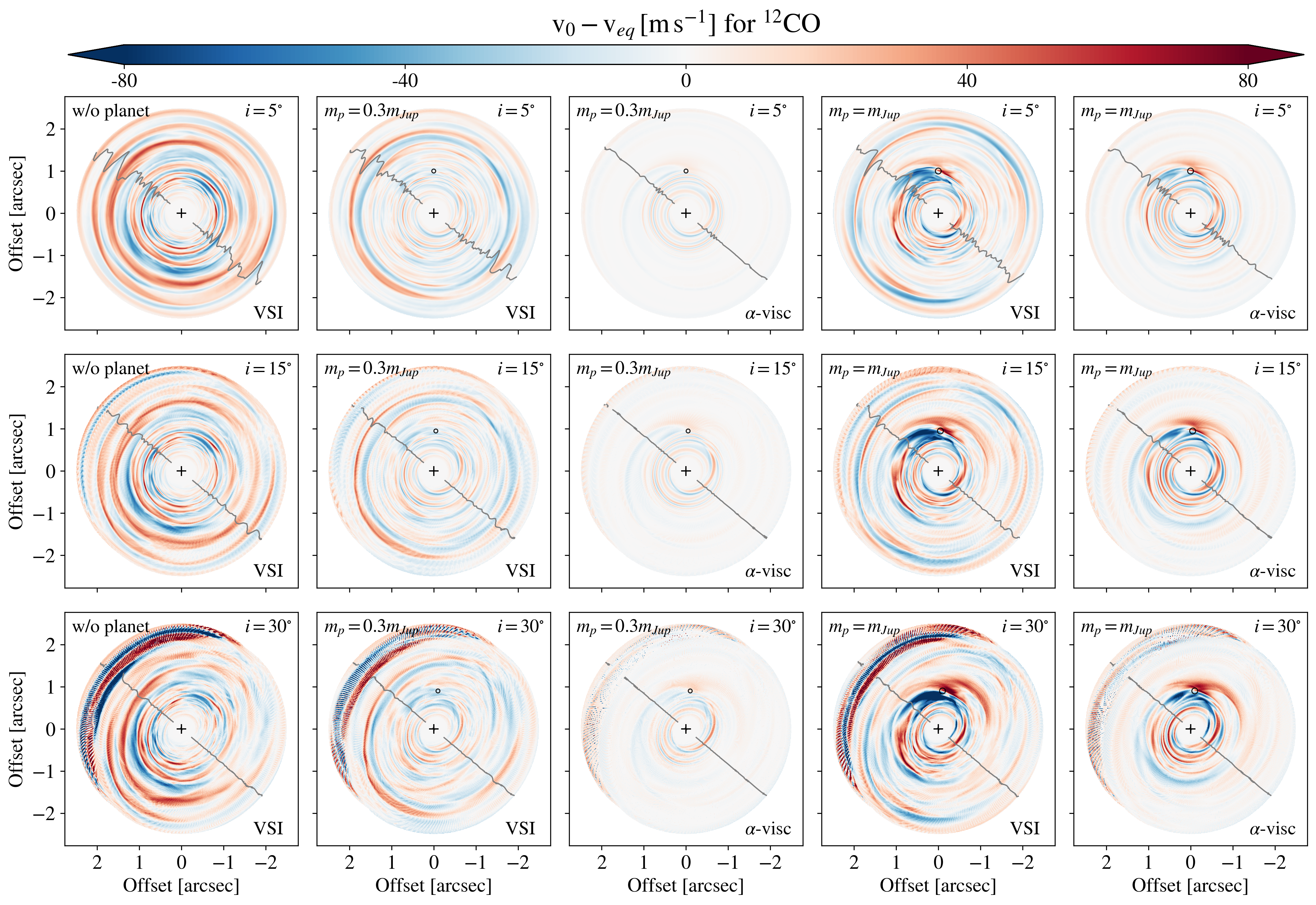}
    \caption{Deviations from Keplerian rotation for the RAW $^{12}$CO$(3-2)$ synthetic predictions of our post-processed simulations. From left to right, are shown: VSI unstable disk without a planet, VSI-unstable disk with a Saturn planet, $\alpha$-viscous disk with a Saturn planet, VSI-unstable disk with a Jupiter planet, and $\alpha$-viscous disk with a Jupiter planet. From top to bottom, predictions for different disk inclinations are presented ($i=[5^{\circ}$, $15^{\circ}$, $30^{\circ}]$). The approximated location of the planets is indicated by black circles with a size of the planets' Hill sphere.}
    \label{fig:residualsraw12co}
\end{figure*}

First, we analyze the kinematic signatures of our disk models in an idealistic case, for images without beam convolution and noise, and with a velocity resolution of 5 m s$^{-1}$. We first study this case in order to have a reference of what would be extracted from our disk model synthetic predictions in an ideal case of unrealistically deep observations and perfect modeling. For this purpose, we extract the deviations from Keplerian rotation in the line-of-sight velocity maps computed from our RAW synthetic radiative transfer images. In order to extract these deviations in the perturbed disk maps, we subtract a second line-of-sight velocity map computed for a disk model following the equilibrium solution used as the initial condition of the simulations. To avoid effects from variations in the traced CO emission layer in the residuals, we only change the model velocities to the equilibrium solutions keeping the same CO number densities and disk temperature structure as the perturbed case. Evidently, this approach is not feasible to apply in real observations; however, as aforementioned, is presented as an ideal picture of the non-Keplerian signatures. We present the $^{12}$CO$(3-2)$ predictions for our set of simulations in Figure \ref{fig:residualsraw12co}, for three different disk inclinations ($5$, $15$ and $30$ degrees). On top of the 2D map residuals from Keplerian rotation a line tracing v$_0=0$ is overlaid, indicating the approximate location of the semi-minor axis and tracing the magnitudes of the distortions created by the non-Keplerian motions in the line-of-sight velocity at the systemic velocity. Again, the disk is oriented in the sky such that its rotation is clockwise and the near-side is at the North-East.\\
In the first column of Figure \ref{fig:residualsraw12co}, we present the kinematic signatures of our VSI-unstable disk model without an embedded planet, showing ring-like residuals tracing the meridional flows produced by the VSI unstable modes, recovering the results presented in \cite{Barraza2021}. 
In the second and third columns, we show the cases with a Saturn-mass planet embedded in a VSI-unstable and an $\alpha$-viscous disk, respectively. We observe that the signatures from the VSI are not seen along the region affected by the planet-induced damping (as discussed in Section \ref{simulationsresults}), while signatures from the VSI are still visible in the outermost parts of the disk. These signatures are mixed with signatures of the spiral driven by the planet via Lindblad resonances, breaking the axisymmetry of the VSI kinematic signatures. The Saturn-mass planet in the $\alpha$-viscous disk produces smooth signatures of Lindblad spirals and spiral wakes around the planet, with weak velocity magnitudes overall.
In columns four and five, we show the cases with an embedded planet with the mass of Jupiter for our VSI-unstable disk and an $\alpha$-viscous disk, respectively. In these scenarios, the massive planet produces a strong signature at its location. Such a feature has been previously denominated 'Doppler-flip' (see, e.g., \citealt{DiskDynamics2020}) and has a significant contribution from the planet's spiral wakes. The planetary spiral wakes produce a super-Keplerian feature outside the planet's radius and a sub-Keplerian feature inside the planet's radius; consequently, creating a dipole pattern.
Additional kinematic signatures are introduced by the planets' inner and outer Lindblad spirals, and the sub- and super-Keplerian rings of gas along the gap and gap edges. Similar to the case for the Saturn-mass planet, for the VSI-unstable disk additional kinematic features are seen in the outermost regions of the disk, in interplay with the planetary spiral arms, which gives a complex kinematic structure with arcs and spiral-like non-Keplerian flows.
From the comparison of the different models varying disk inclination (see Figure \ref{fig:residualsraw12co}), we find that the disk inclination can considerably impact the extracted kinematic signatures. As the disk inclination is increased the planet-driven kinematic signatures are more prominent, while the VSI signatures' velocity magnitude remains fairly constant. These results are expected, since the perturbations in the gas velocity of the disk produced by the planet are strongest in the radial and azimuthal directions, which contribute more to the velocity projected into the line of sight for higher disk inclinations.\\
We also explored the kinematic signatures dependence on the traced disk height by computing the ideal view of the deviations from Keplerian for our $^{13}$CO$(3-2)$ and C$^{18}$O$(3-2)$ predictions, presented in Figures \ref{fig:residualsraw13co} and \ref{fig:residualsrawc18o}, respectively. 
Moving from the most abundant tracer ($^{12}$CO) to the less abundant (C$^{18}$O), deeper layers of the disk are probed, in which the change in the morphology of the kinematic structure for different CO isotopologues strongly depends on the disk inclination. 
In the case of the VSI-unstable disk without embedded planets, the morphology of the non-Keplerian signatures remains fairly consistent for different CO isotopologues, as previously shown in \cite{Barraza2021}. On the contrary, clear changes could be seen among the residuals for the three different CO isotopologues for the models including a massive planet perturbing the disk.
For low disk inclinations, we observe that for less abundant tracers, the planetary-induced non-Keplerian flows are weaker, isolating the VSI operating in the outermost regions of the disk, but with lower velocity magnitudes compared to the $^{12}$CO(3-2) predictions. In addition, the contribution of the Lindblad spirals to the Keplerian model residuals weakens for less abundant tracers independent of disk inclination. For a disk inclination of $30^{\circ}$ the Doppler-flip at the planet location and the super-Keplerian ring at the gap's outer edge remain prominent independent of CO tracer, particularly in the Jupiter case. These features could be explained by the planetary spiral wakes being the strongest dynamical feature at the disk midplane layers, with vigorous azimuthal and radial flows. Similarly, the planet-induced Super-Keplerian ring of gas is fairly independent of disk height (see Figure \ref{fig:azimuthalaverages}). In contrast, VSI flows reach larger velocities at the disk's upper layers, and have a dominant meridional velocity (see Section \ref{simulationsresults}).\\

\subsection{Kinematic Signatures: ALMA simulated observations}\label{simulatedkinematicsignatures}

\begin{figure*}[htp!]
    \centering
    \includegraphics[width=\textwidth]{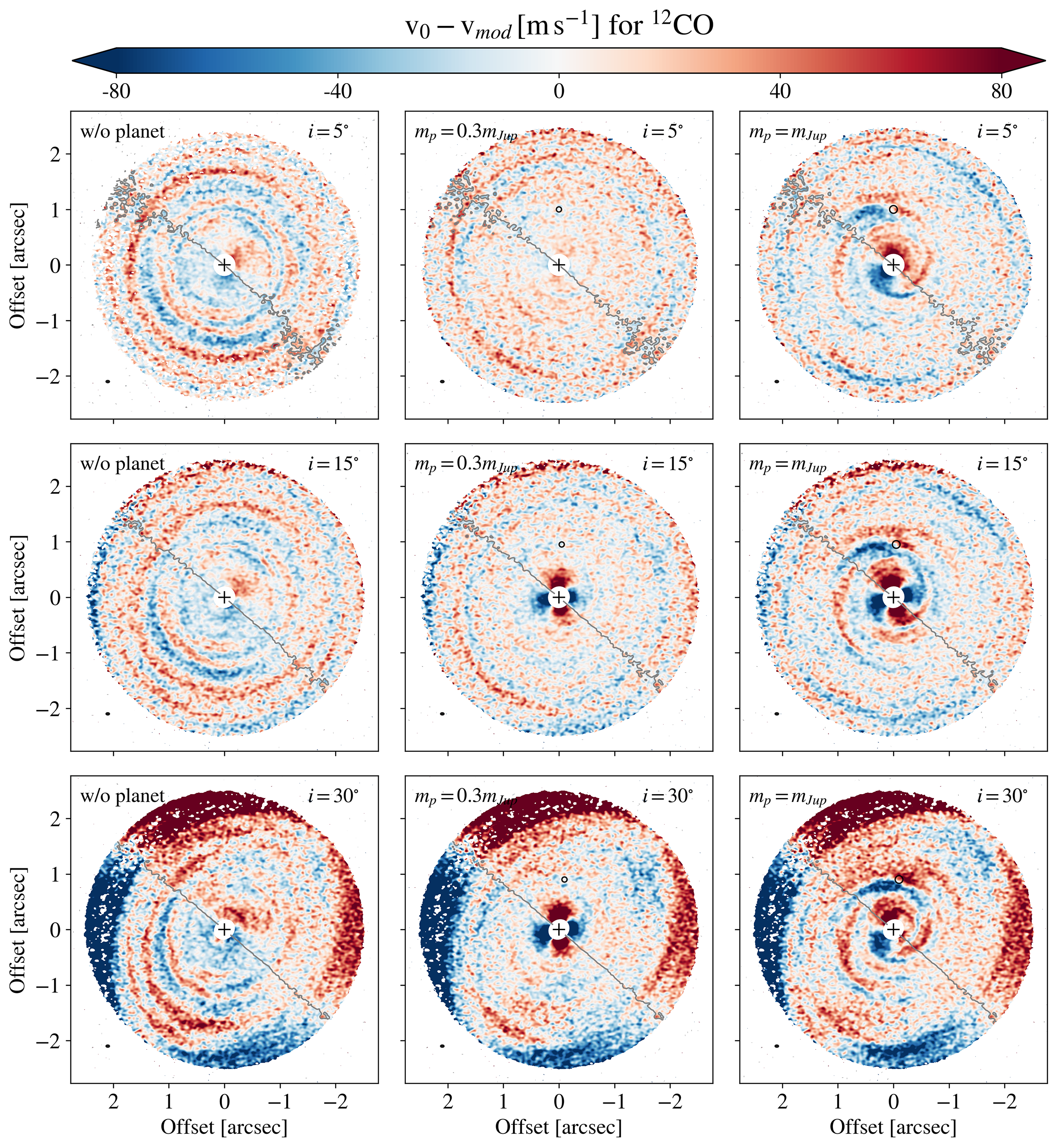}
    \caption{Deviations from Keplerian rotation for $^{12}$CO$(3-2)$ ALMA simulated observations. From left to right: VSI-unstable disk without a planet, VSI-unstable disk with a Saturn mass planet, and VSI-unstable disk with a Jupiter mass planet. From top to bottom: predictions for different disk inclinations ($i=[5^{\circ}$, $15^{\circ}$, $30^{\circ}]$). The approximated location of the planets is indicated by black circles with the size of the planet's Hill radius. The gray solid lines indicate the zero velocity contour in the line-of-sight velocity maps. The disk is oriented so that its rotation in the sky is clockwise. The synthesized beam of $84\times 62$ mas is shown at the left-bottom corner of each panel. The data cubes have a velocity resolution of 100 m s$^{-1}$, with an RMS of 1 mJy beam$^{-1}$ per channel. In all images, regions of low S/N have been clipped.} 
    \label{fig:residualssimobserve}
\end{figure*}

In order to study a more realistic picture of the kinematic signatures that could be observed in VSI-unstable planet-forming disks, we produced simulated ALMA observations for our three VSI-unstable disk models. In Figure \ref{fig:residualssimobserve}, we show the deviations from Keplerian rotation extracted from mock observations of $^{12}$CO$(3-2)$ using \textsc{Eddy} (see Section \ref{kinematictools}), for three different disk inclinations ($i=[5^{\circ}$, $15^{\circ}$, and $30^{\circ}]$). As described in Section \ref{methodssimobserve}, the simulated observations are performed for a combination of ALMA configurations 7 and 4, with a resulting spatial resolution of $84\times62$ mas ($8.4\times 6.2$ au). In terms of spectral resolution, the simulated observations are produced for a velocity resolution of 100 m s$^{-1}$, and a noise level with an RMS of 1 mJy beam$^{-1}$ per channel. These predictions are optimistic and follow the ideal design for kinematic detection of embedded planets \citep{DiskDynamics2020, Pinte2023}. 
While we assume an integration time that gives excellent uv-coverage, to reach the assumed noise levels larger integration times would be needed; for example, at 345 GHz (approximated frequency of the J=3-2 transition of $^{12}$CO), for a column density of water vapor of $\approx 0.9$ mm, such observation would take approximately 40 hours on-source. Nevertheless, these ambitious observations are the goal of the community studying the kinematic structure of protoplanetary disks, which are needed to fully resolve the substructures in the disk gas velocities.\\
The model residuals presented in Fig. \ref{fig:residualssimobserve} show that the deviations from Keplerian rotation induced by the VSI would be observed with clarity only in the case without embedded planets (first column). Arcs of VSI-induced red- and blue-shifted gas are also seen in the outermost regions of the Saturn mass planet case for disk inclinations of 5$^{\circ}$ and 15$^{\circ}$; however, their velocity magnitude is weaker due to the global damping of the VSI induced by the planet, as discussed in \ref{vsiplanetsimulations}. For the highest inclination explored, spiral-like signatures would be observed mixed with VSI arc-like residuals in the outer disk, which would be difficult to differentiate, for example, from signatures of spiral arms triggered by buoyancy resonances \citep{Bae2021}. Nevertheless, we discuss a potential approach to disentangle VSI signatures in Section \ref{vsiplanetsdiscussion}.
In the case of an embedded Jupiter-mass planet, the planet-induced kinematic signatures stand out in the Keplerian model residuals. Important features are a Doppler-flip around the planet and large-scale Lindblad spirals. Unlike the ideal case, super- and sub-Keplerian signatures at the gap edges are weaker, possibly due to the missing modeling of the drop in the emission surface height at the gap region. On top of that, global patterns in the residuals appear due to errors in the model, demonstrating that even with tightly constrained initial values for the free parameters, the fitting can drive errors due to the limitations of the disk model. In particular, a quadrupole pattern is seen in the residual maps near the central region, due to errors in the model center.
This set of simulated ALMA observations suggests that VSI signatures would be easy to identify only in disk regions unperturbed by fairly massive planets, limiting the chances of robust detection of VSI. Also, our results indicate that VSI-turbulent gas motions would not prevent the detection of a Jupiter planet in resolved gas kinematic observations. Finally, the signatures from a VSI-unstable disk with an embedded Saturn-mass planet would be difficult to observe with the current ALMA capabilities and challenging to interpret, so further analysis and observations might be required.\\
Additional substructures could be extracted from exploiting the information of the line profiles, which combined with the line-of-sight velocity maps could potentially disentangle between scenarios. Variations of the line intensity peak and width relative to the disk background could trace deviations of the gas temperature and density for optically thick tracers, possibly tracing spiral arms and gaps produced by embedded planets \citep[e.g.,][]{Bae2021, Izquierdo2021a, Izquierdo2021b}. 
In the case of line peak intensity maps of $^{12}$CO, our set of simulations is not suitable to explore variations on the temperature structure self-consistently, where 3D global simulations including radiative effects are required \citep[e.g.,][]{Bae2021}. In preliminary tests using the temperature structure provided by the thermal Monte Carlo calculations, we obtain spiral-like features, mostly tracing the planet gap region and Lindblad spirals. However, these variations reach values below $1 \%$ relative to the disk background, while in recent observations relative variations up to $5\%$ are found \citep{Teague2019,Teague2021}. Therefore, additional simulations of embedded planets in turbulent protoplanetary disks including radiation-hydrodynamics are needed to robustly explore this observable, and connect it to velocity deviations from Keplerian rotation.\\
In an exploration of line width maps of $^{12}$CO(3-2), we find that variations of this quantity relative to the disk background can trace the planet's gap, and that non-thermal broadening effects are most prominent around the planet's location, consistent with previous studies (e.g., \citealt{Izquierdo2021a}). Moreover, in the residual maps, asymmetries appear inside the gap region for the Jupiter-mass case, also in agreement with previous findings \citep[e.g.,][]{Dong2019, Izquierdo2021a, Izquierdo2021b}. VSI turbulent motions, however, produce arc-like features in the line width maps residuals. These variations are relatively small, reaching values of a few tens of m s$^{-1}$, challenging to extract and interpret \citep[e.g.,][]{Teague2022}. These small line width residuals are consistent with the negligible non-turbulent broadening found in \cite{Barraza2021} for integrated line profiles.
In our line width maps, artifacts from the influence of the back side of the disk when using single-Gaussian fits are seen. Fitting both CO layers (front and back surfaces) is required to overcome such effects \citep[e.g.,][see also Sections \ref{vsiplanetsdiscussion} and \ref{vsiplanetscaveats}]{Izquierdo2021b, Casassus2019}. Careful self-consistent analysis of variations in the peak and width of the CO line will be provided in follow-up studies.

\begin{figure*}[htp!]
    \centering
    \includegraphics[width=\textwidth]{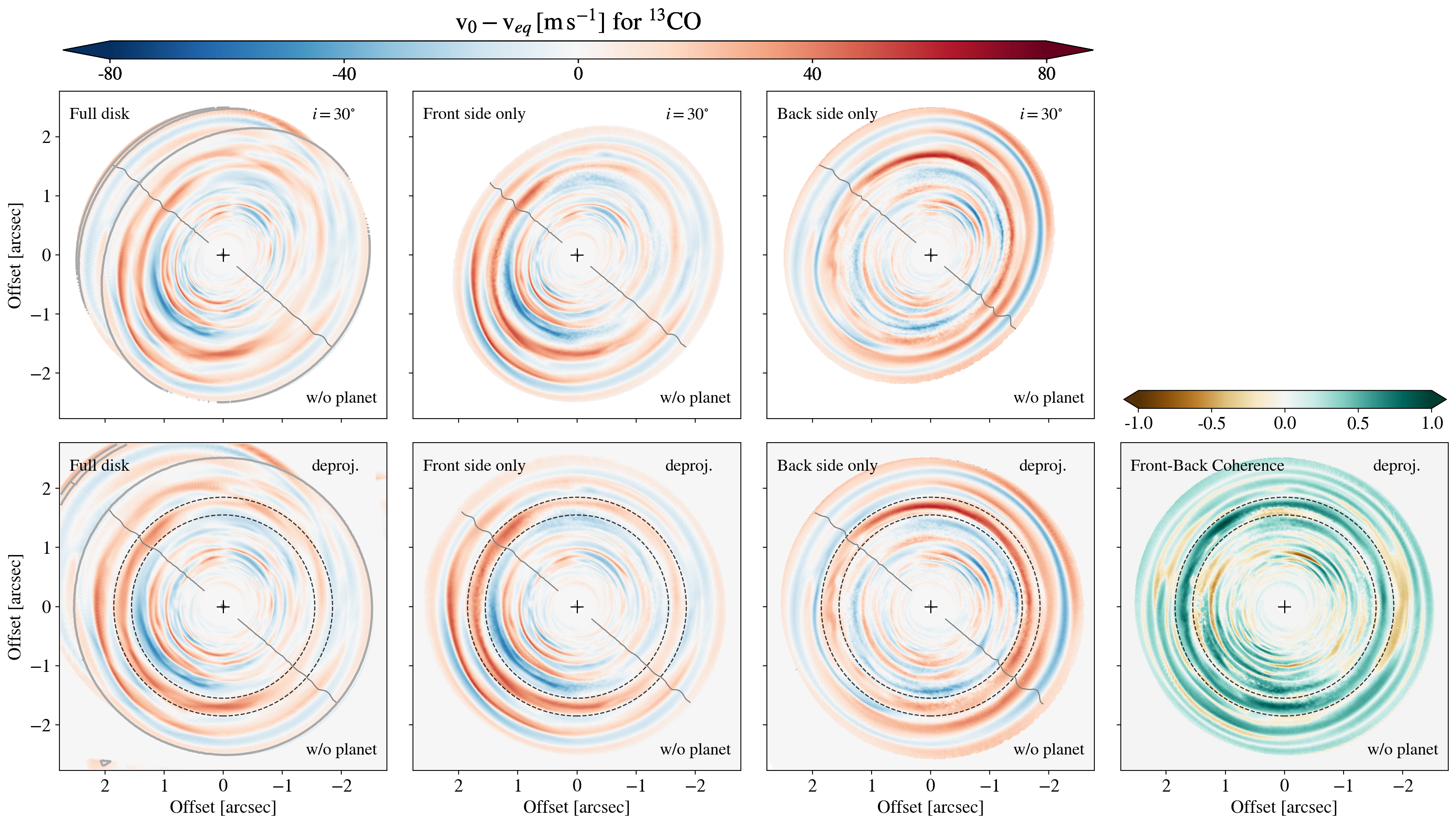}
    \caption{Deviations from Keplerian rotation for $^{13}$CO$(3-2)$ synthetic images of the post-processed simulations, for a VSI unstable disk without embedded planets. In the first row, from left to right, we show models including the full disk, only the disk hemisphere closer to the observer (front side), and the farther hemisphere of the disk (back side), for a disk inclination of 30 degrees. In the second row, a deprojected view of the residual maps (first row) is shown. The fourth panel, shows a normalized square root of the product of the front and back sides deprojected residual maps, times the sign of the initial multiplication of both residuals. A ring of interest is enclosed by black-dashed lines, drawn at 1.55 and 1.85 arcsec. The black solid line along the semi-minor axis in the non-Keplerian residuals shows the line of zero gas velocity projected into the line-of-sight.} 
    \label{fig:residualstwolayersvsi}
\end{figure*}

\section{Discussion}\label{vsiplanetsdiscussion}

\subsection{Can we confirm VSI as the origin of kinematic signatures?}\label{confirmVSI}

To distinguish VSI kinematic signatures from signatures of other mechanisms could be really challenging, due to the possible resemblance of their imprints \citep[see Section 4.2 in][]{Barraza2021}. Moreover, there are physical processes whose simulations suggest could induce similar structures to VSI, yet observational predictions of CO kinematics of these are still lacking. That is the case, for example, for magnetically-driven winds \citep[e.g.,][]{Hu2022}. Therefore, we face the question: If we observe a kinematic structure that matches the expected signature from VSI can we robustly conclude that VSI is operating in the disk?\\
A robust VSI confirmation might be possible by exploiting the information from both disk hemispheres, and invoking the symmetry of VSI flows relative to the midplane layer. As mentioned in Section \ref{vsiplanetsimulations} \citep[see also][]{Bae2021}, flows induced by the VSI have a unique property; With respect to the midplane the VSI flows are: anti-symmetric in the meridional direction, and symmetric in the radial and azimuthal directions. Luckily, it is possible to explore such feature by extracting the kinematic information of both disk hemispheres using observations of CO isotopologues, in which two emitting layers are observed, separated by the colder midplane region (see, e.g., channel maps of the HD 163296 disk presented in \citealt{Pinte2018}).\\
Currently, great efforts are being made to develop techniques to extract both emission CO layers, the front (or upper) and the back (or bottom) surfaces. By fitting a double-Gaussian profile to the collapsed molecular line data at each pixel, the front and back layers have been successfully extracted in the disks HD 135344B \citep{Casassus2021} and HD 163296 \citep{Izquierdo2021b}. However, deeper and higher resolution observations are needed to precisely disentangle both surfaces, which is crucial for kinematic analysis and study of the spatially resolved non-Keplerian flows.\\
In order to demonstrate the possibility of exploiting the VSI symmetry relative to the disk midplane, we ran two models which only take into account one hemisphere of the disk at the time. That is, a model with only the upper hemisphere and a model with only the bottom hemisphere. By computing the expected deviations from Keplerian rotation for RAW $^{13}$CO(3-2) synthetic images of each disk hemisphere, we can compare the front and back $^{13}$CO layer residuals, shown in Figure \ref{fig:residualstwolayersvsi} for a disk inclination of $30^{\circ}$. Under the assumption that we could extract a resolved view of the back CO layer and a good fit of the emission surfaces, VSI quasi-axisymmetric rings of positive and negative residuals are recovered in both, front and back, $^{13}$CO emitting layers. Again, the residuals are dominated by the meridional velocity component, as expected from VSI; therefore, these maps can be interpreted as the column of gas moving to the same vertical direction on both disk hemispheres, a unique feature of VSI. Such symmetry is better seen in a deprojected view of the residual maps, displayed in the bottom row of Fig. \ref{fig:residualstwolayersvsi} (deprojected using \textsc{diskmap\footnote{\url{https://github.com/tomasstolker/diskmap}}} \citealt{Stolker2016}). Additional modulations are seen at the East and West from the disk semi-minor axis for the front and back sides, respectively, coming from azimuthal velocity perturbations moving to opposite directions (i.e., super-Keplerian on one disk hemisphere and sub-Keplerian in the other hemisphere, shown in Figure \ref{fig:azimuthalaverages} and \ref{fig:simradialprofiles}).\\
A visualization of the meridional flows moving towards the same direction is shown in a 'coherence' \footnote{We apply the term 'coherence' to refer to a gas flow moving in the meridional direction as a coherent structure through both disk hemispheres at a particular radius. In the opposite case, the flow would be divided into two distinct structures, both moving toward (or away from) the midplane.} map (last panel of Figure \ref{fig:residualstwolayersvsi}). We define the coherence of the deprojected line-of-sight velocities from the upper and bottom layers as: 

\begin{equation}
    \textrm{v$_{0,\rm coherence}$} = \sgn \left( \textrm{v$_{0,\rm upper}^{\rm deproj.}$}\times \textrm{v$_{0,\rm lower}^{\rm deproj.}$}\right) \sqrt{ \left| \textrm{v$_{0,\rm upper}^{deproj.}$}\times \textrm{v$_{0,\rm lower}^{\rm deproj.}$} \right| },
\end{equation}

\noindent where v$_{0,\rm upper}^{\rm deproj.}$, and v$_{0,\rm lower}^{\rm deproj.}$ are the deprojected line-of-sight velocities of the upper and bottom layers, respectively. The coherence map shown in Figure \ref{fig:residualstwolayersvsi} displays the normalized coherence, where rings of positive values reveal coherence of the vertical flow. Particularly clear coherence is seen in the ring of interest enclosed by black dashed lines, with both sides moving away from the observer. When compared to the cases of VSI-unstable disks with embedded massive planets (shown in Figures \ref{fig:residualstwolayerssaturn} and \ref{fig:residualstwolayersjupiter}), we observe that VSI quasi-axisymmetric rings of positive coherence are only present in the outermost regions of the disks with planets; once again demonstrating the planet-induced damping of the VSI. Moreover, such a coherence map also highlights the perturbations induced by the Jupiter-mass planet around its location (last panel of Fig. \ref{fig:residualstwolayersjupiter}), with potential use to point towards localized massive planet signatures.\\
Nevertheless, exploiting both CO emission layers of the planet-forming disk is extremely challenging. On top of the high resolution required, mm-sized dust grains at the disk midplane can block parts of the emission from the back side layer, complicating even more its study. Despite our simplifications, it has been demonstrated that exploring both hemispheres of the disk in CO kinematic observations is possible. Thus, future deep high-resolution molecular line observations can confirm the symmetry of the VSI flows with respect to the midplane, a robust proof of the VSI operating in planet-forming disks.\\
Current studies show that meridional flows induced by alternative mechanisms are directly correlated to the formation of a deep gap in the gas. Therefore, the gas would flow from the disk surface towards the gas-depleted region. Such a correlation is not typical for VSI signatures. As seen in Figure \ref{fig:azimuthalaverages}, the direction of the meridional flows is not tightly correlated with the density perturbations. Thus, quasi-axisymmetric meridional flows in a region without deep gaps in the gas would be consistent with VSI-induced kinematic signatures. Finally, in the opposite case, where deep gaps in the gas are correlated with the meridional flows, VSI is unlikely to be the origin, favoring massive planets or non-ideal magneto-hydrodynamical effects.\\

\subsection{Caveats}\label{vsiplanetscaveats}

In the following, we will discuss the limitations of our approach.
\indent First, the assumption of a locally isothermal equation of state in our hydrodynamical simulations is certainly a simplification. Under the assumption of a fast cooling disk (cooling timescales substantially shorter than the local orbital timescale), needed for the VSI to operate, this is still a valid approximation. However, the locally isothermal approach, which translates to an instantaneous cooling ($t_{\rm cool=0}$), is an extreme limit, which leads to vigorous VSI \citep[e.g., ][]{Flores2020, Cui2022}. Even in rapidly-cooling disks (e.g., $t_{\rm cool} \sim 0.01 t_{\rm orb}$), the VSI will be slightly damped compared to a locally isothermal disk \citep{Manger2021}. Nevertheless, simulations relaxing the isothermal assumption for fast cooling disks lead to similar VSI gas dynamics \citep[e.g.,][]{Stoll2014, Flock2017}, and planet-disk interactions \citep{Ziampras2023} to the obtained in our simulations. 
On the other hand, simulations including regions with long cooling timescales, that do not fulfill the requirements for the VSI to operate, will result in a confined VSI-active layer \citep{Pfeil2021, Fukahara2023} or almost complete suppression of the VSI, directly affecting the observability of VSI-signatures in gas kinematics. The cooling properties of the disk, thus, the ability of the disk to sustain VSI, is primarily regulated by the amount and properties of small dust grains in the disk \citep{Malygin2017}. Conditions of inefficient cooling in the outer disk can originate from infrequent dust and gas collisions near the disk atmosphere \citep[e.g., ][]{Pfeil2021, Bae2021}, and/or an overall reduced dust-to-gas ratio of the small dust grains as a result of dust evolution \citep{Lin2015,Fukahara2021,Dullemond2022,Pfeil2023}. Including regions with longer cooling times would also lead to additional planet-induced signatures from spirals generated via buoyancy resonances \citep{Bae2021}.

Second, the assumed thermal structure of the disk in the hydrodynamical simulations is not consistent with the theoretical predictions nor the observed structure from recent ALMA observations, in which a vertical gradient of the temperature is constrained. Such structure can modify the development of the VSI and planet-driven structures. Nevertheless, previous work has shown that in simulations including both, disk thermal structure and radiation-hydrodynamics, the VSI still operates vigorously in the fast cooling regions of the disk \citep{Flock2020}. An additional set of simulations including planets in a VSI unstable disks with vertical temperature structure has still to be performed.

Third, our radiative transfer models assume constant n(CO)/n(H$_2$) through the disk. Such an assumption is likely not to hold if the influence of VSI and planet-driven large-scale flows on the disk chemistry is taken into account. Mixing of material in the radial and vertical directions would modify the spatial distribution of molecules, therefore, changing the emitting surface, and abundance radial profiles of CO isotopologues \citep[e.g.,][]{Semenov2006, Semenov2011}. Moreover, a planet-carved gap can significantly affect the density and thermal structure of the disk, altering the abundance of CO isotopologues around the gap location. These thermo-chemical effects are important to be included in future studies, as they would impact the observability of kinematic signatures in the circumstellar disk.

Finally, in our work, a single Gaussian function is used to extract the information from the simulated ALMA observations. Such an approach could result in a line fitting merging information from the back and front emitting layers of CO, producing a velocity structure that does not fully reflect the disk gas dynamics. These effects are relevant for disks with intermediate inclinations, predominant in regions with lower gas densities (e.g., planet-carved gaps), while minimal along the disk semi-major and semi-minor axis. In our models, the obtained morphology of the velocity residuals from Keplerian rotation reflects the expected morphology from the front CO layer, as demonstrated in Figure \ref{fig:residualstwolayersvsi}.
The implementation of routines to fit both emission layers at the same time is ideal for extracting the true velocity structures from the disk \citep[e.g.,][]{Casassus2019, Casassus2021, Izquierdo2021a, Izquierdo2021b}. Moreover, as discussed in Section  \ref{confirmVSI}, such an approach could help to disentangle VSI signatures from planet-induced signatures. Exploring the effects of applying an improved line fitting procedure in our models is left for follow-up work.\\

\section{Summary and Conclusions}\label{conclusions}

In this paper, we presented a comprehensive study of the gas dynamics and kinematic signatures of planet-forming disks unstable to the vertical shear instability (VSI). Particularly, we explored the interplay between the VSI and structures induced by an embedded massive planet, and their resulting signatures observable in CO rotational line observations with ALMA.
We performed this study by running global 3D hydrodynamical simulations, post-processed with radiative transfer calculations, to finally simulate mock ALMA observations. 
Specifically, we studied the effects on the disk dynamical structure of single planets with the mass of Saturn and Jupiter, and their imprints on the observable deviations from Keplerian rotation.\\
We found that the presence of fairly massive planets embedded in the disk substantially affects the gas velocity structure produced by the VSI, damping the VSI unstable modes in the regions where the planets significantly modify the structure of the disk. Further, the damping is stronger by increasing the planet's mass, and is most effective in a region near the midplane layer of the disk.\\
The effect of the planets on the VSI motions significantly alters the kinematic signatures. The observable kinematic signatures of the VSI are globally weakened, and only clearly visible tens of au radially outward from the planets' location. The VSI adds fine structure to the planet-induced kinematic signatures, with a complex interplay in the Saturn-mass case. For the case of an embedded Jupiter-mass planet, the planet-induced signatures dominate the kinematic structure of the disk, showing a clear Doppler-flip at the planet's location and spiral arms in the residuals from a Keplerian model.\\
Furthermore, we compare simulations of VSI-unstable disk and simulations following a constant $\alpha$ viscosity prescription. This direct comparison highlights the predictions of the additional kinematic signatures produced by the VSI compared to the standard $\alpha$ viscous case.
The more complex kinematic structure, found for the Saturn-mass planet case, showing a mixture of VSI modes and planet-induced spirals, might impede the identification of the planet and VSI in ALMA observations. 
Thus, simultaneous modeling of different CO isotopologues might be needed for robust planet detection, where the best strategy to isolate the planet signatures is to observe closer to the midplane of moderately inclined disks.\\
Finally, we test an approach to confirm the presence of VSI motions in future high-resolution ALMA observations, by detecting the coherence of the perturbations with respect to the disk midplane. Such an approach is promising for revealing the VSI operating in disks.\\
We conclude that the best chance to detect clear VSI signatures is to look for disk regions distant from observed deep continuum or molecular gas gaps, where the VSI-induced perturbations might still be active far from the influence of putative massive planets. In addition, exploring the flows' symmetries with respect to the disk midplane is a pathway to confirm VSI signatures in future CO rotational line observations.\\
We highlight the potential of directly comparing deep ALMA CO observations with theoretical predictions of kinematic signatures. Robust interpretations could reveal the presence of embedded massive planets, signatures of disk instabilities, and constrain disk physical properties. 
In the near future, upgrades planned for the ALMA interferometer infrastructure \citep[ALMA2030;][]{Carpenter2023} will significantly increase the sensitivity for line emission observations. This technological advance will allow a deeper study of planet-forming disks kinematics, revealing the fine structure of gas flows, probing regions closer to the disk midplane, and possibly resolving the circumplanetary region of embedded massive planets; thus, potentially revealing a comprehensive picture of planet-disk interactions in turbulent protoplanetary disks.

\begin{acknowledgements}

 We thank the anonymous referee for providing constructive comments on the manuscript. We thank the developers and contributors of the codes and software used throughout this work, including the developers of the Python packages \textsc{Numpy} \citep{Harris2020}, \textsc{Scipy} \citep{Virtanen2020}, \textsc{Astropy} \citep{Astropy2013} and \textsc{Matplotlib} \cite{Hunter2007}. M.B. thanks R. Teague and L. Flores-Rivera for providing constructive feedback on figures, and S. Andrews and N. Kurtovic for their advice in the use of \textsc{CASA} \texttt{tclean}. M.B. thanks the exoALMA collaboration for fruitful discussions on protoplanetary disk kinematics. M.B. and M.F. acknowledge support from the European Research Council (ERC), under the European Union’s Horizon 2020 research and innovation program (grant agreement No. 757957). T.H. acknowledge support from the European Research Council under the Horizon 2020 Framework Program via the ERC Advanced Grant Origins 83 24 28. The set of numerical simulations presented was conducted on the COBRA supercomputer, hosted by the Max Planck Computing and Data Facility (MPCDF).

\end{acknowledgements}

\bibliographystyle{aa} 
\bibliography{references} 

\begin{thebibliography}{113}
\expandafter\ifx\csname natexlab\endcsname\relax\def\natexlab#1{#1}\fi

\bibitem[{{Andrews} {et~al.}(2018){Andrews}, {Huang}, {P{\'e}rez}, {Isella}, {Dullemond}, {Kurtovic}, {Guzm{\'a}n}, {Carpenter}, {Wilner}, {Zhang}, {Zhu}, {Birnstiel}, {Bai}, {Benisty}, {Hughes}, {{\"O}berg}, \& {Ricci}}]{Andrews2018}
{Andrews}, S.~M., {Huang}, J., {P{\'e}rez}, L.~M., {et~al.} 2018, \apjl, 869, L41

\bibitem[{{Asensio-Torres} {et~al.}(2021){Asensio-Torres}, {Henning}, {Cantalloube}, {Pinilla}, {Mesa}, {Garufi}, {Jorquera}, {Gratton}, {Chauvin}, {Szul{\'a}gyi}, {van Boekel}, {Dong}, {Marleau}, {Benisty}, {Villenave}, {Bergez-Casalou}, {Desgrange}, {Janson}, {Keppler}, {Langlois}, {M{\'e}nard}, {Rickman}, {Stolker}, {Feldt}, {Fusco}, {Gluck}, {Pavlov}, \& {Ramos}}]{AsensioTorres2021}
{Asensio-Torres}, R., {Henning}, T., {Cantalloube}, F., {et~al.} 2021, \aap, 652, A101

\bibitem[{{Astropy Collaboration} {et~al.}(2013){Astropy Collaboration}, {Robitaille}, {Tollerud}, {Greenfield}, {Droettboom}, {Bray}, {Aldcroft}, {Davis}, {Ginsburg}, {Price-Whelan}, {Kerzendorf}, {Conley}, {Crighton}, {Barbary}, {Muna}, {Ferguson}, {Grollier}, {Parikh}, {Nair}, {Unther}, {Deil}, {Woillez}, {Conseil}, {Kramer}, {Turner}, {Singer}, {Fox}, {Weaver}, {Zabalza}, {Edwards}, {Azalee Bostroem}, {Burke}, {Casey}, {Crawford}, {Dencheva}, {Ely}, {Jenness}, {Labrie}, {Lim}, {Pierfederici}, {Pontzen}, {Ptak}, {Refsdal}, {Servillat}, \& {Streicher}}]{Astropy2013}
{Astropy Collaboration}, {Robitaille}, T.~P., {Tollerud}, E.~J., {et~al.} 2013, \aap, 558, A33

\bibitem[{{Bae} {et~al.}(2023){Bae}, {Isella}, {Zhu}, {Martin}, {Okuzumi}, \& {Suriano}}]{Bae2023}
{Bae}, J., {Isella}, A., {Zhu}, Z., {et~al.} 2023, in Astronomical Society of the Pacific Conference Series, Vol. 534, Astronomical Society of the Pacific Conference Series, ed. S.~{Inutsuka}, Y.~{Aikawa}, T.~{Muto}, K.~{Tomida}, \& M.~{Tamura}, 423

\bibitem[{{Bae} {et~al.}(2021){Bae}, {Teague}, \& {Zhu}}]{Bae2021}
{Bae}, J., {Teague}, R., \& {Zhu}, Z. 2021, \apj, 912, 56

\bibitem[{{Bae} \& {Zhu}(2018{\natexlab{a}})}]{Bae2018}
{Bae}, J. \& {Zhu}, Z. 2018{\natexlab{a}}, \apj, 859, 118

\bibitem[{{Bae} \& {Zhu}(2018{\natexlab{b}})}]{Bae2018b}
{Bae}, J. \& {Zhu}, Z. 2018{\natexlab{b}}, \apj, 859, 119

\bibitem[{{Bae} {et~al.}(2019){Bae}, {Zhu}, {Baruteau}, {Benisty}, {Dullemond}, {Facchini}, {Isella}, {Keppler}, {P{\'e}rez}, \& {Teague}}]{Bae2019}
{Bae}, J., {Zhu}, Z., {Baruteau}, C., {et~al.} 2019, \apjl, 884, L41

\bibitem[{{Balbus} \& {Hawley}(1991)}]{Balbus1991}
{Balbus}, S.~A. \& {Hawley}, J.~F. 1991, \apj, 376, 214

\bibitem[{{Balbus} \& {Hawley}(1998)}]{Balbus1998}
{Balbus}, S.~A. \& {Hawley}, J.~F. 1998, Reviews of Modern Physics, 70, 1

\bibitem[{{Barraza-Alfaro} {et~al.}(2021){Barraza-Alfaro}, {Flock}, {Marino}, \& {P{\'e}rez}}]{Barraza2021}
{Barraza-Alfaro}, M., {Flock}, M., {Marino}, S., \& {P{\'e}rez}, S. 2021, \aap, 653, A113

\bibitem[{{Baruteau} {et~al.}(2019){Baruteau}, {Barraza}, {P{\'e}rez}, {Casassus}, {Dong}, {Lyra}, {Marino}, {Christiaens}, {Zhu}, {Carmona}, {Debras}, \& {Alarcon}}]{Baruteau2019}
{Baruteau}, C., {Barraza}, M., {P{\'e}rez}, S., {et~al.} 2019, \mnras, 486, 304

\bibitem[{{Benisty} {et~al.}(2021){Benisty}, {Bae}, {Facchini}, {Keppler}, {Teague}, {Isella}, {Kurtovic}, {P{\'e}rez}, {Sierra}, {Andrews}, {Carpenter}, {Czekala}, {Dominik}, {Henning}, {Menard}, {Pinilla}, \& {Zurlo}}]{Benisty2021}
{Benisty}, M., {Bae}, J., {Facchini}, S., {et~al.} 2021, \apjl, 916, L2

\bibitem[{{Benisty} {et~al.}(2023){Benisty}, {Dominik}, {Follette}, {Garufi}, {Ginski}, {Hashimoto}, {Keppler}, {Kley}, \& {Monnier}}]{Benisty2023}
{Benisty}, M., {Dominik}, C., {Follette}, K., {et~al.} 2023, in Astronomical Society of the Pacific Conference Series, Vol. 534, Protostars and Planets VII, ed. S.~{Inutsuka}, Y.~{Aikawa}, T.~{Muto}, K.~{Tomida}, \& M.~{Tamura}, 605

\bibitem[{{Bergez-Casalou} {et~al.}(2020){Bergez-Casalou}, {Bitsch}, {Pierens}, {Crida}, \& {Raymond}}]{Bergez2020}
{Bergez-Casalou}, C., {Bitsch}, B., {Pierens}, A., {Crida}, A., \& {Raymond}, S.~N. 2020, \aap, 643, A133

\bibitem[{{Bi} {et~al.}(2021){Bi}, {Lin}, \& {Dong}}]{Bi2021}
{Bi}, J., {Lin}, M.-K., \& {Dong}, R. 2021, \apj, 912, 107

\bibitem[{{Binkert} {et~al.}(2021){Binkert}, {Szul{\'a}gyi}, \& {Birnstiel}}]{Binkert2021}
{Binkert}, F., {Szul{\'a}gyi}, J., \& {Birnstiel}, T. 2021, \mnras, 506, 5969

\bibitem[{{Bohren} \& {Huffman}(1983)}]{Bohren1983}
{Bohren}, C.~F. \& {Huffman}, D.~R. 1983, {Absorption and scattering of light by small particles}

\bibitem[{{Brown-Sevilla} {et~al.}(2021){Brown-Sevilla}, {Keppler}, {Barraza-Alfaro}, {Melon Fuksman}, {Kurtovic}, {Pinilla}, {Feldt}, {Brandner}, {Ginski}, {Henning}, {Klahr}, {Asensio-Torres}, {Cantalloube}, {Garufi}, {van Holstein}, {Langlois}, {M{\'e}nard}, {Rickman}, {Benisty}, {Chauvin}, {Zurlo}, {Weber}, {Pavlov}, {Ramos}, {Rochat}, \& {Roelfsema}}]{Brown2021}
{Brown-Sevilla}, S.~B., {Keppler}, M., {Barraza-Alfaro}, M., {et~al.} 2021, \aap, 654, A35

\bibitem[{{Carpenter} {et~al.}(2023){Carpenter}, {Brogan}, {Iono}, \& {Mroczkowski}}]{Carpenter2023}
{Carpenter}, J., {Brogan}, C., {Iono}, D., \& {Mroczkowski}, T. 2023, in Physics and Chemistry of Star Formation: The Dynamical ISM Across Time and Spatial Scales, 304

\bibitem[{{Casassus} {et~al.}(2021){Casassus}, {Christiaens}, {C{\'a}rcamo}, {P{\'e}rez}, {Weber}, {Ercolano}, {van der Marel}, {Pinte}, {Dong}, {Baruteau}, {Cieza}, {van Dishoeck}, {Jordan}, {Price}, {Absil}, {Arce-Tord}, {Faramaz}, {Flores}, \& {Reggiani}}]{Casassus2021}
{Casassus}, S., {Christiaens}, V., {C{\'a}rcamo}, M., {et~al.} 2021, \mnras, 507, 3789

\bibitem[{{Casassus} \& {P{\'e}rez}(2019)}]{Casassus2019}
{Casassus}, S. \& {P{\'e}rez}, S. 2019, \apjl, 883, L41

\bibitem[{{Crida} \& {Morbidelli}(2007)}]{Crida2007}
{Crida}, A. \& {Morbidelli}, A. 2007, \mnras, 377, 1324

\bibitem[{{Cui} \& {Bai}(2022)}]{Cui2022}
{Cui}, C. \& {Bai}, X.-N. 2022, \mnras, 516, 4660

\bibitem[{{D'Angelo} {et~al.}(2002){D'Angelo}, {Henning}, \& {Kley}}]{DAngelo2002}
{D'Angelo}, G., {Henning}, T., \& {Kley}, W. 2002, \aap, 385, 647

\bibitem[{{D'Angelo} {et~al.}(2003){D'Angelo}, {Henning}, \& {Kley}}]{DAngelo2003}
{D'Angelo}, G., {Henning}, T., \& {Kley}, W. 2003, \apj, 599, 548

\bibitem[{{de Val-Borro} {et~al.}(2007){de Val-Borro}, {Artymowicz}, {D'Angelo}, \& {Peplinski}}]{deValBorro2007}
{de Val-Borro}, M., {Artymowicz}, P., {D'Angelo}, G., \& {Peplinski}, A. 2007, \aap, 471, 1043

\bibitem[{{de Val-Borro} {et~al.}(2006){de Val-Borro}, {Edgar}, {Artymowicz}, {Ciecielag}, {Cresswell}, {D'Angelo}, {Delgado-Donate}, {Dirksen}, {Fromang}, {Gawryszczak}, {Klahr}, {Kley}, {Lyra}, {Masset}, {Mellema}, {Nelson}, {Paardekooper}, {Peplinski}, {Pierens}, {Plewa}, {Rice}, {Sch{\"a}fer}, \& {Speith}}]{deValBorro2006}
{de Val-Borro}, M., {Edgar}, R.~G., {Artymowicz}, P., {et~al.} 2006, \mnras, 370, 529

\bibitem[{{Disk Dynamics Collaboration} {et~al.}(2020){Disk Dynamics Collaboration}, {Armitage}, {Bae}, {Benisty}, {Bergin}, {Casassus}, {Czekala}, {Facchini}, {Fung}, {Hall}, {Ilee}, {Keppler}, {Kuznetsova}, {Le Gal}, {Loomis}, {Lyra}, {Manger}, {Perez}, {Pinte}, {Price}, {Rosotti}, {Szulagyi}, {Schwarz}, {Simon}, {Teague}, \& {Zhang}}]{DiskDynamics2020}
{Disk Dynamics Collaboration}, {Armitage}, P.~J., {Bae}, J., {et~al.} 2020, arXiv e-prints, arXiv:2009.04345

\bibitem[{{Dominik} {et~al.}(2021){Dominik}, {Min}, \& {Tazaki}}]{Dominik2021}
{Dominik}, C., {Min}, M., \& {Tazaki}, R. 2021, {OpTool: Command-line driven tool for creating complex dust opacities}, Astrophysics Source Code Library, record ascl:2104.010

\bibitem[{{Dong} {et~al.}(2019){Dong}, {Liu}, \& {Fung}}]{Dong2019}
{Dong}, R., {Liu}, S.-Y., \& {Fung}, J. 2019, \apj, 870, 72

\bibitem[{{Dr{\k{a}}{\.z}kowska} {et~al.}(2023){Dr{\k{a}}{\.z}kowska}, {Bitsch}, {Lambrechts}, {Mulders}, {Harsono}, {Vazan}, {Liu}, {Ormel}, {Kretke}, \& {Morbidelli}}]{Drazkowska2023}
{Dr{\k{a}}{\.z}kowska}, J., {Bitsch}, B., {Lambrechts}, M., {et~al.} 2023, in Astronomical Society of the Pacific Conference Series, Vol. 534, Protostars and Planets VII, ed. S.~{Inutsuka}, Y.~{Aikawa}, T.~{Muto}, K.~{Tomida}, \& M.~{Tamura}, 717

\bibitem[{{Dullemond} {et~al.}(2012){Dullemond}, {Juhasz}, {Pohl}, {Sereshti}, {Shetty}, {Peters}, {Commercon}, \& {Flock}}]{Dullemond2012}
{Dullemond}, C.~P., {Juhasz}, A., {Pohl}, A., {et~al.} 2012, {RADMC-3D: A multi-purpose radiative transfer tool}, Astrophysics Source Code Library, record ascl:1202.015

\bibitem[{{Dullemond} {et~al.}(2022){Dullemond}, {Ziampras}, {Ostertag}, \& {Dominik}}]{Dullemond2022}
{Dullemond}, C.~P., {Ziampras}, A., {Ostertag}, D., \& {Dominik}, C. 2022, \aap, 668, A105

\bibitem[{{Flaherty} {et~al.}(2020){Flaherty}, {Hughes}, {Simon}, {Qi}, {Bai}, {Bulatek}, {Andrews}, {Wilner}, \& {K{\'o}sp{\'a}l}}]{Flaherty2020}
{Flaherty}, K., {Hughes}, A.~M., {Simon}, J.~B., {et~al.} 2020, \apj, 895, 109

\bibitem[{{Flaherty} {et~al.}(2018){Flaherty}, {Hughes}, {Teague}, {Simon}, {Andrews}, \& {Wilner}}]{Flaherty2018}
{Flaherty}, K.~M., {Hughes}, A.~M., {Teague}, R., {et~al.} 2018, \apj, 856, 117

\bibitem[{{Flock} {et~al.}(2017){Flock}, {Nelson}, {Turner}, {Bertrang}, {Carrasco-Gonz{\'a}lez}, {Henning}, {Lyra}, \& {Teague}}]{Flock2017}
{Flock}, M., {Nelson}, R.~P., {Turner}, N.~J., {et~al.} 2017, \apj, 850, 131

\bibitem[{{Flock} {et~al.}(2020){Flock}, {Turner}, {Nelson}, {Lyra}, {Manger}, \& {Klahr}}]{Flock2020}
{Flock}, M., {Turner}, N.~J., {Nelson}, R.~P., {et~al.} 2020, \apj, 897, 155

\bibitem[{{Flores-Rivera} {et~al.}(2020){Flores-Rivera}, {Flock}, \& {Nakatani}}]{Flores2020}
{Flores-Rivera}, L., {Flock}, M., \& {Nakatani}, R. 2020, \aap, 644, A50

\bibitem[{{Fukuhara} {et~al.}(2021){Fukuhara}, {Okuzumi}, \& {Ono}}]{Fukahara2021}
{Fukuhara}, Y., {Okuzumi}, S., \& {Ono}, T. 2021, \apj, 914, 132

\bibitem[{{Fukuhara} {et~al.}(2023){Fukuhara}, {Okuzumi}, \& {Ono}}]{Fukahara2023}
{Fukuhara}, Y., {Okuzumi}, S., \& {Ono}, T. 2023, \pasj, 75, 233

\bibitem[{Fung \& Chiang(2016)}]{Fung2016}
Fung, J. \& Chiang, E. 2016, The Astrophysical Journal, 832, 105

\bibitem[{{Galloway-Sprietsma} {et~al.}(2023){Galloway-Sprietsma}, {Bae}, {Teague}, {Benisty}, {Facchini}, {Aikawa}, {Alarc{\'o}n}, {Andrews}, {Bergin}, {Cataldi}, {Cleeves}, {Czekala}, {Guzm{\'a}n}, {Huang}, {Law}, {Le Gal}, {Liu}, {Long}, {M{\'e}nard}, {{\"O}berg}, {Walsh}, \& {Wilner}}]{Galloway2023}
{Galloway-Sprietsma}, M., {Bae}, J., {Teague}, R., {et~al.} 2023, \apj, 950, 147

\bibitem[{{Gammie}(1996)}]{Gammie1996}
{Gammie}, C.~F. 1996, \apj, 457, 355

\bibitem[{{Garg} {et~al.}(2022){Garg}, {Pinte}, {Hammond}, {Teague}, {Hilder}, {Price}, {Calcino}, {Christiaens}, \& {Poblete}}]{Garg2022}
{Garg}, H., {Pinte}, C., {Hammond}, I., {et~al.} 2022, \mnras [\eprint[arXiv]{2210.10248}]

\bibitem[{{Garufi} {et~al.}(2018){Garufi}, {Benisty}, {Pinilla}, {Tazzari}, {Dominik}, {Ginski}, {Henning}, {Kral}, {Langlois}, {M{\'e}nard}, {Stolker}, {Szulagyi}, {Villenave}, \& {van der Plas}}]{Garufi2018}
{Garufi}, A., {Benisty}, M., {Pinilla}, P., {et~al.} 2018, \aap, 620, A94

\bibitem[{{Goodman} \& {Rafikov}(2001)}]{Goodman2001}
{Goodman}, J. \& {Rafikov}, R.~R. 2001, \apj, 552, 793

\bibitem[{{Gressel} {et~al.}(2013){Gressel}, {Nelson}, {Turner}, \& {Ziegler}}]{Gressel2013}
{Gressel}, O., {Nelson}, R.~P., {Turner}, N.~J., \& {Ziegler}, U. 2013, \apj, 779, 59

\bibitem[{{Hall} {et~al.}(2020){Hall}, {Dong}, {Teague}, {Terry}, {Pinte}, {Paneque-Carre{\~n}o}, {Veronesi}, {Alexander}, \& {Lodato}}]{Hall2020}
{Hall}, C., {Dong}, R., {Teague}, R., {et~al.} 2020, \apj, 904, 148

\bibitem[{{Hammer} \& {Lin}(2023)}]{Hammer2023}
{Hammer}, M. \& {Lin}, M.-K. 2023, \mnras, 525, 123

\bibitem[{Harris {et~al.}(2020)Harris, Millman, van~der Walt, Gommers, Virtanen, Cournapeau, Wieser, Taylor, Berg, Smith, Kern, Picus, Hoyer, van Kerkwijk, Brett, Haldane, del R{\'{i}}o, Wiebe, Peterson, G{\'{e}}rard-Marchant, Sheppard, Reddy, Weckesser, Abbasi, Gohlke, \& Oliphant}]{Harris2020}
Harris, C.~R., Millman, K.~J., van~der Walt, S.~J., {et~al.} 2020, Nature, 585, 357

\bibitem[{{Hawley} \& {Balbus}(1991)}]{Hawley1991}
{Hawley}, J.~F. \& {Balbus}, S.~A. 1991, \apj, 376, 223

\bibitem[{{Hu} {et~al.}(2022){Hu}, {Li}, {Zhu}, \& {Yang}}]{Hu2022}
{Hu}, X., {Li}, Z.-Y., {Zhu}, Z., \& {Yang}, C.-C. 2022, \mnras, 516, 2006

\bibitem[{{Hunter}(2007)}]{Hunter2007}
{Hunter}, J.~D. 2007, Computing in Science and Engineering, 9, 90

\bibitem[{{Izquierdo} {et~al.}(2022){Izquierdo}, {Facchini}, {Rosotti}, {van Dishoeck}, \& {Testi}}]{Izquierdo2021b}
{Izquierdo}, A.~F., {Facchini}, S., {Rosotti}, G.~P., {van Dishoeck}, E.~F., \& {Testi}, L. 2022, \apj, 928, 2

\bibitem[{{Izquierdo} {et~al.}(2021){Izquierdo}, {Testi}, {Facchini}, {Rosotti}, \& {van Dishoeck}}]{Izquierdo2021a}
{Izquierdo}, A.~F., {Testi}, L., {Facchini}, S., {Rosotti}, G.~P., \& {van Dishoeck}, E.~F. 2021, \aap, 650, A179

\bibitem[{{Kepley} {et~al.}(2020){Kepley}, {Tsutsumi}, {Brogan}, {Indebetouw}, {Yoon}, {Mason}, \& {Donovan Meyer}}]{Kepley2020}
{Kepley}, A.~A., {Tsutsumi}, T., {Brogan}, C.~L., {et~al.} 2020, \pasp, 132, 024505

\bibitem[{{Keppler} {et~al.}(2018){Keppler}, {Benisty}, {M{\"u}ller}, {Henning}, {van Boekel}, {Cantalloube}, {Ginski}, {van Holstein}, {Maire}, {Pohl}, {Samland}, {Avenhaus}, {Baudino}, {Boccaletti}, {de Boer}, {Bonnefoy}, {Chauvin}, {Desidera}, {Langlois}, {Lazzoni}, {Marleau}, {Mordasini}, {Pawellek}, {Stolker}, {Vigan}, {Zurlo}, {Birnstiel}, {Brandner}, {Feldt}, {Flock}, {Girard}, {Gratton}, {Hagelberg}, {Isella}, {Janson}, {Juhasz}, {Kemmer}, {Kral}, {Lagrange}, {Launhardt}, {Matter}, {M{\'e}nard}, {Milli}, {Molli{\`e}re}, {Olofsson}, {P{\'e}rez}, {Pinilla}, {Pinte}, {Quanz}, {Schmidt}, {Udry}, {Wahhaj}, {Williams}, {Buenzli}, {Cudel}, {Dominik}, {Galicher}, {Kasper}, {Lannier}, {Mesa}, {Mouillet}, {Peretti}, {Perrot}, {Salter}, {Sissa}, {Wildi}, {Abe}, {Antichi}, {Augereau}, {Baruffolo}, {Baudoz}, {Bazzon}, {Beuzit}, {Blanchard}, {Brems}, {Buey}, {De Caprio}, {Carbillet}, {Carle}, {Cascone}, {Cheetham}, {Claudi}, {Costille}, {Delboulb{\'e}}, {Dohlen}, {Fantinel}, {Feautrier}, {Fusco}, {Giro}, {Gluck},
  {Gry}, {Hubin}, {Hugot}, {Jaquet}, {Le Mignant}, {Llored}, {Madec}, {Magnard}, {Martinez}, {Maurel}, {Meyer}, {M{\"o}ller-Nilsson}, {Moulin}, {Mugnier}, {Orign{\'e}}, {Pavlov}, {Perret}, {Petit}, {Pragt}, {Puget}, {Rabou}, {Ramos}, {Rigal}, {Rochat}, {Roelfsema}, {Rousset}, {Roux}, {Salasnich}, {Sauvage}, {Sevin}, {Soenke}, {Stadler}, {Suarez}, {Turatto}, \& {Weber}}]{Keppler2018}
{Keppler}, M., {Benisty}, M., {M{\"u}ller}, A., {et~al.} 2018, \aap, 617, A44

\bibitem[{{Keppler} {et~al.}(2019){Keppler}, {Teague}, {Bae}, {Benisty}, {Henning}, {van Boekel}, {Chapillon}, {Pinilla}, {Williams}, {Bertrang}, {Facchini}, {Flock}, {Ginski}, {Juhasz}, {Klahr}, {Liu}, {M{\"u}ller}, {P{\'e}rez}, {Pohl}, {Rosotti}, {Samland}, \& {Semenov}}]{Keppler2019}
{Keppler}, M., {Teague}, R., {Bae}, J., {et~al.} 2019, \aap, 625, A118

\bibitem[{{Klahr} \& {Hubbard}(2014)}]{Klahr2014}
{Klahr}, H. \& {Hubbard}, A. 2014, \apj, 788, 21

\bibitem[{{Kley} {et~al.}(2001){Kley}, {D'Angelo}, \& {Henning}}]{Kley2001}
{Kley}, W., {D'Angelo}, G., \& {Henning}, T. 2001, \apj, 547, 457

\bibitem[{{Lehmann} \& {Lin}(2022)}]{Lehmann2022}
{Lehmann}, M. \& {Lin}, M.~K. 2022, \aap, 658, A156

\bibitem[{{Lenz} {et~al.}(2020){Lenz}, {Klahr}, {Birnstiel}, {Kretke}, \& {Stammler}}]{Lenz2020}
{Lenz}, C.~T., {Klahr}, H., {Birnstiel}, T., {Kretke}, K., \& {Stammler}, S. 2020, \aap, 640, A61

\bibitem[{{Lesur} {et~al.}(2023){Lesur}, {Flock}, {Ercolano}, {Lin}, {Yang}, {Barranco}, {Benitez-Llambay}, {Goodman}, {Johansen}, {Klahr}, {Laibe}, {Lyra}, {Marcus}, {Nelson}, {Squire}, {Simon}, {Turner}, {Umurhan}, \& {Youdin}}]{Lesur2023}
{Lesur}, G., {Flock}, M., {Ercolano}, B., {et~al.} 2023, in Astronomical Society of the Pacific Conference Series, Vol. 534, Astronomical Society of the Pacific Conference Series, ed. S.~{Inutsuka}, Y.~{Aikawa}, T.~{Muto}, K.~{Tomida}, \& M.~{Tamura}, 465

\bibitem[{{Li} {et~al.}(2000){Li}, {Finn}, {Lovelace}, \& {Colgate}}]{Li2000}
{Li}, H., {Finn}, J.~M., {Lovelace}, R.~V.~E., \& {Colgate}, S.~A. 2000, \apj, 533, 1023

\bibitem[{{Lin} \& {Papaloizou}(1993)}]{Lin1993}
{Lin}, D.~N.~C. \& {Papaloizou}, J.~C.~B. 1993, in Protostars and Planets III, ed. E.~H. {Levy} \& J.~I. {Lunine}, 749

\bibitem[{{Lin} \& {Youdin}(2015)}]{Lin2015}
{Lin}, M.-K. \& {Youdin}, A.~N. 2015, \apj, 811, 17

\bibitem[{{Lovelace} {et~al.}(1999){Lovelace}, {Li}, {Colgate}, \& {Nelson}}]{Lovelace1999}
{Lovelace}, R.~V.~E., {Li}, H., {Colgate}, S.~A., \& {Nelson}, A.~F. 1999, \apj, 513, 805

\bibitem[{{Lyra} \& {Umurhan}(2019)}]{Lyra2019}
{Lyra}, W. \& {Umurhan}, O.~M. 2019, \pasp, 131, 072001

\bibitem[{{Malygin} {et~al.}(2017){Malygin}, {Klahr}, {Semenov}, {Henning}, \& {Dullemond}}]{Malygin2017}
{Malygin}, M.~G., {Klahr}, H., {Semenov}, D., {Henning}, T., \& {Dullemond}, C.~P. 2017, \aap, 605, A30

\bibitem[{{Manger} \& {Klahr}(2018)}]{Manger2018}
{Manger}, N. \& {Klahr}, H. 2018, \mnras, 480, 2125

\bibitem[{{Manger} {et~al.}(2021){Manger}, {Pfeil}, \& {Klahr}}]{Manger2021}
{Manger}, N., {Pfeil}, T., \& {Klahr}, H. 2021, \mnras, 508, 5402

\bibitem[{{Marcus} {et~al.}(2015){Marcus}, {Pei}, {Jiang}, {Barranco}, {Hassanzadeh}, \& {Lecoanet}}]{Marcus2015}
{Marcus}, P.~S., {Pei}, S., {Jiang}, C.-H., {et~al.} 2015, \apj, 808, 87

\bibitem[{{McMullin} {et~al.}(2007){McMullin}, {Waters}, {Schiebel}, {Young}, \& {Golap}}]{McMullin2007}
{McMullin}, J.~P., {Waters}, B., {Schiebel}, D., {Young}, W., \& {Golap}, K. 2007, in Astronomical Society of the Pacific Conference Series, Vol. 376, Astronomical Data Analysis Software and Systems XVI, ed. R.~A. {Shaw}, F.~{Hill}, \& D.~J. {Bell}, 127

\bibitem[{{Mignone} {et~al.}(2007){Mignone}, {Bodo}, {Massaglia}, {Matsakos}, {Tesileanu}, {Zanni}, \& {Ferrari}}]{Mignone2007}
{Mignone}, A., {Bodo}, G., {Massaglia}, S., {et~al.} 2007, \apjs, 170, 228

\bibitem[{{Nelson} {et~al.}(2013){Nelson}, {Gressel}, \& {Umurhan}}]{Nelson2013}
{Nelson}, R.~P., {Gressel}, O., \& {Umurhan}, O.~M. 2013, \mnras, 435, 2610

\bibitem[{{{\"O}berg} {et~al.}(2021){{\"O}berg}, {Guzm{\'a}n}, {Walsh}, {Aikawa}, {Bergin}, {Law}, {Loomis}, {Alarc{\'o}n}, {Andrews}, {Bae}, {Bergner}, {Boehler}, {Booth}, {Bosman}, {Calahan}, {Cataldi}, {Cleeves}, {Czekala}, {Furuya}, {Huang}, {Ilee}, {Kurtovic}, {Le Gal}, {Liu}, {Long}, {M{\'e}nard}, {Nomura}, {P{\'e}rez}, {Qi}, {Schwarz}, {Sierra}, {Teague}, {Tsukagoshi}, {Yamato}, {van't Hoff}, {Waggoner}, {Wilner}, \& {Zhang}}]{Oberg2021}
{{\"O}berg}, K.~I., {Guzm{\'a}n}, V.~V., {Walsh}, C., {et~al.} 2021, \apjs, 257, 1

\bibitem[{{Paardekooper} {et~al.}(2023){Paardekooper}, {Dong}, {Duffell}, {Fung}, {Masset}, {Ogilvie}, \& {Tanaka}}]{Paardekooper2023}
{Paardekooper}, S., {Dong}, R., {Duffell}, P., {et~al.} 2023, in Astronomical Society of the Pacific Conference Series, Vol. 534, Astronomical Society of the Pacific Conference Series, ed. S.~{Inutsuka}, Y.~{Aikawa}, T.~{Muto}, K.~{Tomida}, \& M.~{Tamura}, 685

\bibitem[{{P{\'e}rez} {et~al.}(2018){P{\'e}rez}, {Casassus}, \& {Ben{\'\i}tez-Llambay}}]{Perez2018}
{P{\'e}rez}, S., {Casassus}, S., \& {Ben{\'\i}tez-Llambay}, P. 2018, \mnras, 480, L12

\bibitem[{{Perez} {et~al.}(2015){Perez}, {Dunhill}, {Casassus}, {Roman}, {Szul{\'a}gyi}, {Flores}, {Marino}, \& {Montesinos}}]{Perez2015}
{Perez}, S., {Dunhill}, A., {Casassus}, S., {et~al.} 2015, \apjl, 811, L5

\bibitem[{{Pfeil} {et~al.}(2023){Pfeil}, {Birnstiel}, \& {Klahr}}]{Pfeil2023}
{Pfeil}, T., {Birnstiel}, T., \& {Klahr}, H. 2023, arXiv e-prints, arXiv:2310.07332

\bibitem[{{Pfeil} \& {Klahr}(2019)}]{Pfeil2019}
{Pfeil}, T. \& {Klahr}, H. 2019, \apj, 871, 150

\bibitem[{{Pfeil} \& {Klahr}(2021)}]{Pfeil2021}
{Pfeil}, T. \& {Klahr}, H. 2021, \apj, 915, 130

\bibitem[{{Pinte} {et~al.}(2018){Pinte}, {Price}, {M{\'e}nard}, {Duch{\^e}ne}, {Dent}, {Hill}, {de Gregorio-Monsalvo}, {Hales}, \& {Mentiplay}}]{Pinte2018}
{Pinte}, C., {Price}, D.~J., {M{\'e}nard}, F., {et~al.} 2018, \apjl, 860, L13

\bibitem[{{Pinte} {et~al.}(2023){Pinte}, {Teague}, {Flaherty}, {Hall}, {Facchini}, \& {Casassus}}]{Pinte2023}
{Pinte}, C., {Teague}, R., {Flaherty}, K., {et~al.} 2023, in Astronomical Society of the Pacific Conference Series, Vol. 534, Astronomical Society of the Pacific Conference Series, ed. S.~{Inutsuka}, Y.~{Aikawa}, T.~{Muto}, K.~{Tomida}, \& M.~{Tamura}, 645

\bibitem[{{Rabago} \& {Zhu}(2021)}]{Rabago2021}
{Rabago}, I. \& {Zhu}, Z. 2021, \mnras, 502, 5325

\bibitem[{{Sch{\"o}ier} {et~al.}(2005){Sch{\"o}ier}, {van der Tak}, {van Dishoeck}, \& {Black}}]{Schoier2005}
{Sch{\"o}ier}, F.~L., {van der Tak}, F.~F.~S., {van Dishoeck}, E.~F., \& {Black}, J.~H. 2005, \aap, 432, 369

\bibitem[{{Segura-Cox} {et~al.}(2020){Segura-Cox}, {Schmiedeke}, {Pineda}, {Stephens}, {Fern{\'a}ndez-L{\'o}pez}, {Looney}, {Caselli}, {Li}, {Mundy}, {Kwon}, \& {Harris}}]{Segura-Cox2020}
{Segura-Cox}, D.~M., {Schmiedeke}, A., {Pineda}, J.~E., {et~al.} 2020, \nat, 586, 228

\bibitem[{{Semenov} \& {Wiebe}(2011)}]{Semenov2011}
{Semenov}, D. \& {Wiebe}, D. 2011, \apjs, 196, 25

\bibitem[{{Semenov} {et~al.}(2006){Semenov}, {Wiebe}, \& {Henning}}]{Semenov2006}
{Semenov}, D., {Wiebe}, D., \& {Henning}, T. 2006, \apjl, 647, L57

\bibitem[{{Shakura} \& {Sunyaev}(1973)}]{Shakura1973}
{Shakura}, N.~I. \& {Sunyaev}, R.~A. 1973, \aap, 500, 33

\bibitem[{{Stadler} {et~al.}(2023){Stadler}, {Benisty}, {Izquierdo}, {Facchini}, {Teague}, {Kurtovic}, {Pinilla}, {Bae}, {Ansdell}, {Loomis}, {Mayama}, {Perez}, \& {Testi}}]{Stadler2023}
{Stadler}, J., {Benisty}, M., {Izquierdo}, A., {et~al.} 2023, \aap, 670, L1

\bibitem[{{Stolker} {et~al.}(2016){Stolker}, {Dominik}, {Min}, {Garufi}, {Mulders}, \& {Avenhaus}}]{Stolker2016}
{Stolker}, T., {Dominik}, C., {Min}, M., {et~al.} 2016, \aap, 596, A70

\bibitem[{{Stoll} \& {Kley}(2014)}]{Stoll2014}
{Stoll}, M. H.~R. \& {Kley}, W. 2014, \aap, 572, A77

\bibitem[{{Stoll} {et~al.}(2017{\natexlab{a}}){Stoll}, {Kley}, \& {Picogna}}]{Stoll2017a}
{Stoll}, M. H.~R., {Kley}, W., \& {Picogna}, G. 2017{\natexlab{a}}, \aap, 599, L6

\bibitem[{{Stoll} {et~al.}(2017{\natexlab{b}}){Stoll}, {Picogna}, \& {Kley}}]{Stoll2017}
{Stoll}, M. H.~R., {Picogna}, G., \& {Kley}, W. 2017{\natexlab{b}}, \aap, 604, A28

\bibitem[{{Svanberg} {et~al.}(2022){Svanberg}, {Cui}, \& {Latter}}]{Svanberg2022}
{Svanberg}, E., {Cui}, C., \& {Latter}, H.~N. 2022, \mnras, 514, 4581

\bibitem[{{Szul{\'a}gyi} {et~al.}(2014){Szul{\'a}gyi}, {Morbidelli}, {Crida}, \& {Masset}}]{Szulagyi2014}
{Szul{\'a}gyi}, J., {Morbidelli}, A., {Crida}, A., \& {Masset}, F. 2014, \apj, 782, 65

\bibitem[{Teague(2019)}]{eddy}
Teague, R. 2019, The Journal of Open Source Software, 4, 1220

\bibitem[{{Teague}(2019)}]{Teague2019b}
{Teague}, R. 2019, Research Notes of the American Astronomical Society, 3, 74

\bibitem[{{Teague} {et~al.}(2021){Teague}, {Bae}, {Aikawa}, {Andrews}, {Bergin}, {Bergner}, {Boehler}, {Booth}, {Bosman}, {Cataldi}, {Czekala}, {Guzm{\'a}n}, {Huang}, {Ilee}, {Law}, {Le Gal}, {Long}, {Loomis}, {M{\'e}nard}, {{\"O}berg}, {P{\'e}rez}, {Schwarz}, {Sierra}, {Walsh}, {Wilner}, {Yamato}, \& {Zhang}}]{Teague2021}
{Teague}, R., {Bae}, J., {Aikawa}, Y., {et~al.} 2021, \apjs, 257, 18

\bibitem[{{Teague} {et~al.}(2022){Teague}, {Bae}, {Andrews}, {Benisty}, {Bergin}, {Facchini}, {Huang}, {Longarini}, \& {Wilner}}]{Teague2022}
{Teague}, R., {Bae}, J., {Andrews}, S.~M., {et~al.} 2022, \apj, 936, 163

\bibitem[{{Teague} {et~al.}(2019{\natexlab{a}}){Teague}, {Bae}, \& {Bergin}}]{Teague2019c}
{Teague}, R., {Bae}, J., \& {Bergin}, E.~A. 2019{\natexlab{a}}, \nat, 574, 378

\bibitem[{{Teague} {et~al.}(2019{\natexlab{b}}){Teague}, {Bae}, {Huang}, \& {Bergin}}]{Teague2019}
{Teague}, R., {Bae}, J., {Huang}, J., \& {Bergin}, E.~A. 2019{\natexlab{b}}, \apjl, 884, L56

\bibitem[{{Teague} \& {Foreman-Mackey}(2018)}]{Teague2018c}
{Teague}, R. \& {Foreman-Mackey}, D. 2018, Research Notes of the American Astronomical Society, 2, 173

\bibitem[{{Teague} {et~al.}(2016){Teague}, {Guilloteau}, {Semenov}, {Henning}, {Dutrey}, {Pi{\'e}tu}, {Birnstiel}, {Chapillon}, {Hollenbach}, \& {Gorti}}]{Teague2016}
{Teague}, R., {Guilloteau}, S., {Semenov}, D., {et~al.} 2016, \aap, 592, A49

\bibitem[{{Teague} {et~al.}(2018){Teague}, {Henning}, {Guilloteau}, {Bergin}, {Semenov}, {Dutrey}, {Flock}, {Gorti}, \& {Birnstiel}}]{Teague2018}
{Teague}, R., {Henning}, T., {Guilloteau}, S., {et~al.} 2018, \apj, 864, 133

\bibitem[{{van Boekel} {et~al.}(2017){van Boekel}, {Henning}, {Menu}, {de Boer}, {Langlois}, {M{\"u}ller}, {Avenhaus}, {Boccaletti}, {Schmid}, {Thalmann}, {Benisty}, {Dominik}, {Ginski}, {Girard}, {Gisler}, {Lobo Gomes}, {Menard}, {Min}, {Pavlov}, {Pohl}, {Quanz}, {Rabou}, {Roelfsema}, {Sauvage}, {Teague}, {Wildi}, \& {Zurlo}}]{vanBoekel2017}
{van Boekel}, R., {Henning}, T., {Menu}, J., {et~al.} 2017, \apj, 837, 132

\bibitem[{{Villenave} {et~al.}(2020){Villenave}, {M{\'e}nard}, {Dent}, {Duch{\^e}ne}, {Stapelfeldt}, {Benisty}, {Boehler}, {van der Plas}, {Pinte}, {Telkamp}, {Wolff}, {Flores}, {Lesur}, {Louvet}, {Riols}, {Dougados}, {Williams}, \& {Padgett}}]{Villenave2020}
{Villenave}, M., {M{\'e}nard}, F., {Dent}, W.~R.~F., {et~al.} 2020, \aap, 642, A164

\bibitem[{{Villenave} {et~al.}(2022){Villenave}, {Stapelfeldt}, {Duch{\^e}ne}, {M{\'e}nard}, {Lambrechts}, {Sierra}, {Flores}, {Dent}, {Wolff}, {Ribas}, {Benisty}, {Cuello}, \& {Pinte}}]{Villenave2022}
{Villenave}, M., {Stapelfeldt}, K.~R., {Duch{\^e}ne}, G., {et~al.} 2022, \apj, 930, 11

\bibitem[{Virtanen {et~al.}(2020)Virtanen, Gommers, Oliphant, Haberland, Reddy, Cournapeau, Burovski, Peterson, Weckesser, Bright, {van der Walt}, Brett, Wilson, Millman, Mayorov, Nelson, Jones, Kern, Larson, Carey, Polat, Feng, Moore, {VanderPlas}, Laxalde, Perktold, Cimrman, Henriksen, Quintero, Harris, Archibald, Ribeiro, Pedregosa, {van Mulbregt}, \& {SciPy 1.0 Contributors}}]{Virtanen2020}
Virtanen, P., Gommers, R., Oliphant, T.~E., {et~al.} 2020, Nature Methods, 17, 261

\bibitem[{{W{\"o}lfer} {et~al.}(2023){W{\"o}lfer}, {Facchini}, {van der Marel}, {van Dishoeck}, {Benisty}, {Bohn}, {Francis}, {Izquierdo}, \& {Teague}}]{Wolfer2023}
{W{\"o}lfer}, L., {Facchini}, S., {van der Marel}, N., {et~al.} 2023, \aap, 670, A154

\bibitem[{{Ziampras} {et~al.}(2023){Ziampras}, {Kley}, \& {Nelson}}]{Ziampras2023}
{Ziampras}, A., {Kley}, W., \& {Nelson}, R.~P. 2023, \aap, 670, A135

\end{thebibliography}

\appendix
\onecolumn
\section{Additional Figures}
\begin{figure*}[htp]
    \centering
    \includegraphics[width=0.8\textwidth]{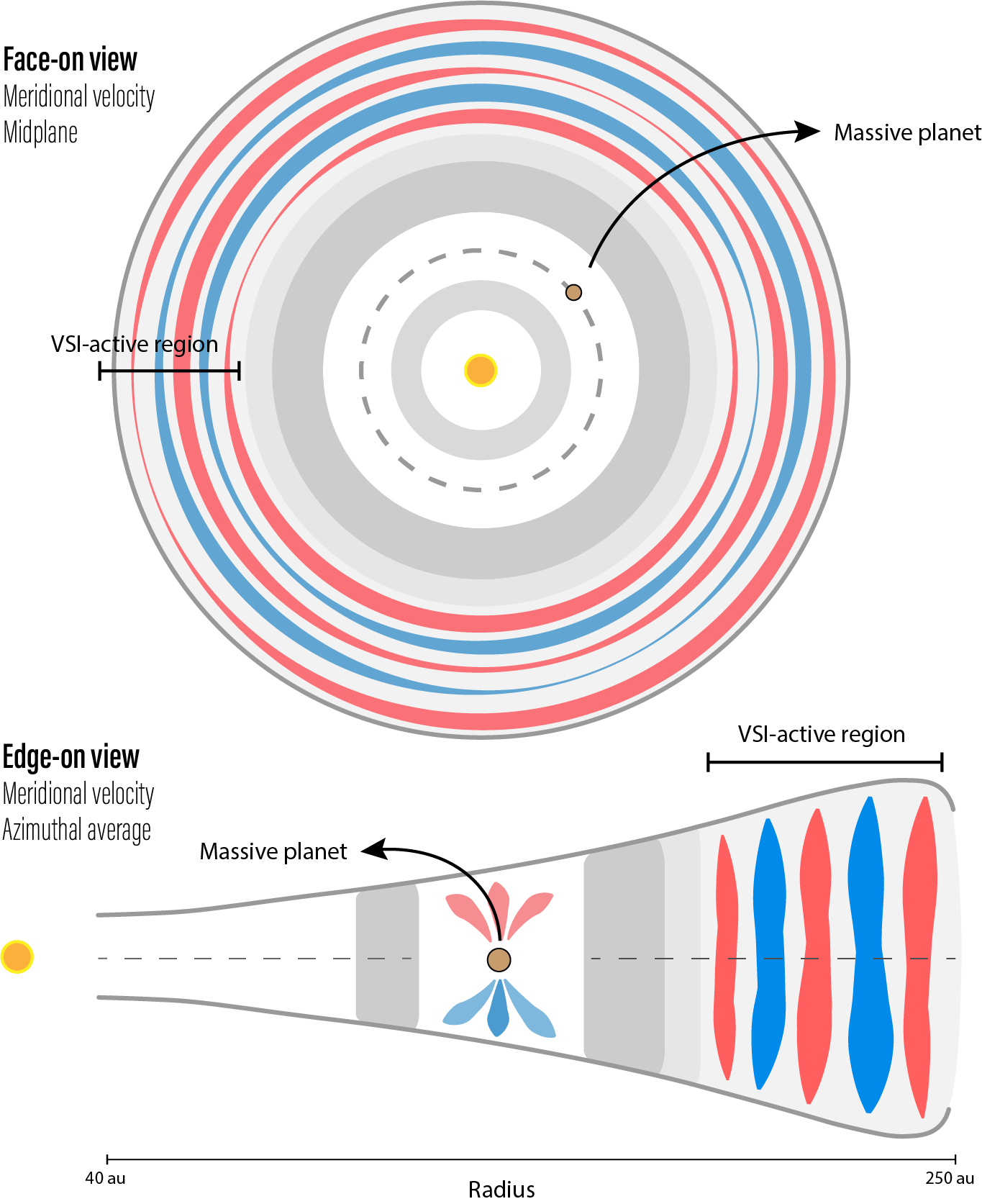}
    \caption{Schematic of the expected meridional velocity ($v_{\theta}$) from a VSI-unstable disk with an embedded massive planet. Top: Face-on view, red (blue) indicates gas moving toward (away from) the disk midplane. Bottom: Edge-on view (radius vs. height) of the predicted azimuthally-averaged, red (blue) indicates gas moving downwards (upwards). The gray colors indicate the gas density.}
    \label{fig:vsiplanetsketch}
\end{figure*}

\begin{figure*}[htp]
    \centering
    \includegraphics[angle=0,width=\linewidth]{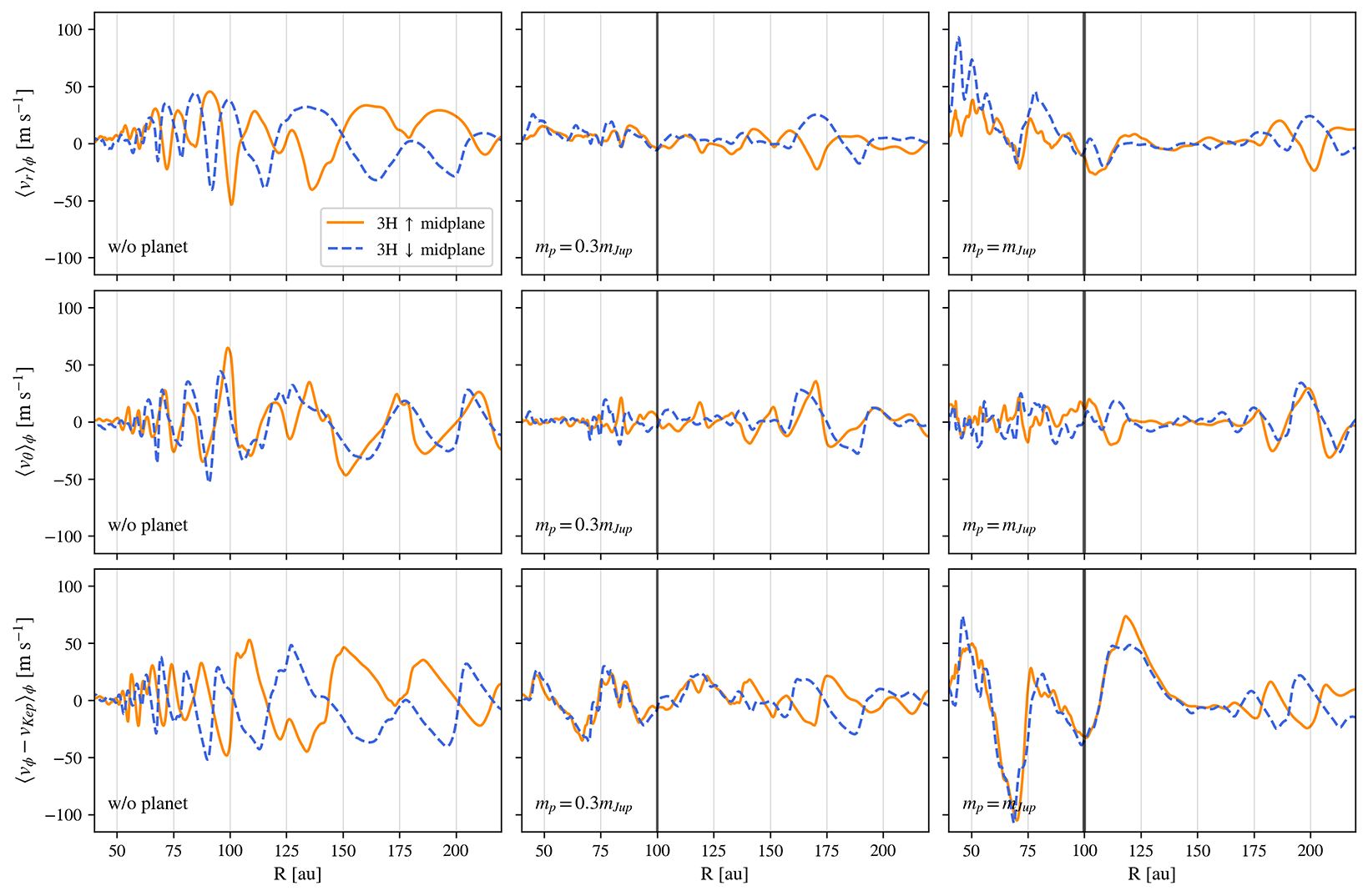}
    \caption{Velocity radial profiles for the VSI-unstable disk simulations without a planet (first column), with an embedded Saturn-mass planet (second column), and with an embedded Jupiter-mass planet (third column). The values displayed correspond to 3 pressure scale heights from the disk midplane for both disk hemispheres, shown by blue-dashed and orange-solid lines. The radial location of the planets is indicated with a vertical black solid line.
    } 
    \label{fig:simradialprofiles}
\end{figure*}

\begin{figure*}[htp]
    \centering
    \includegraphics[width=\textwidth]{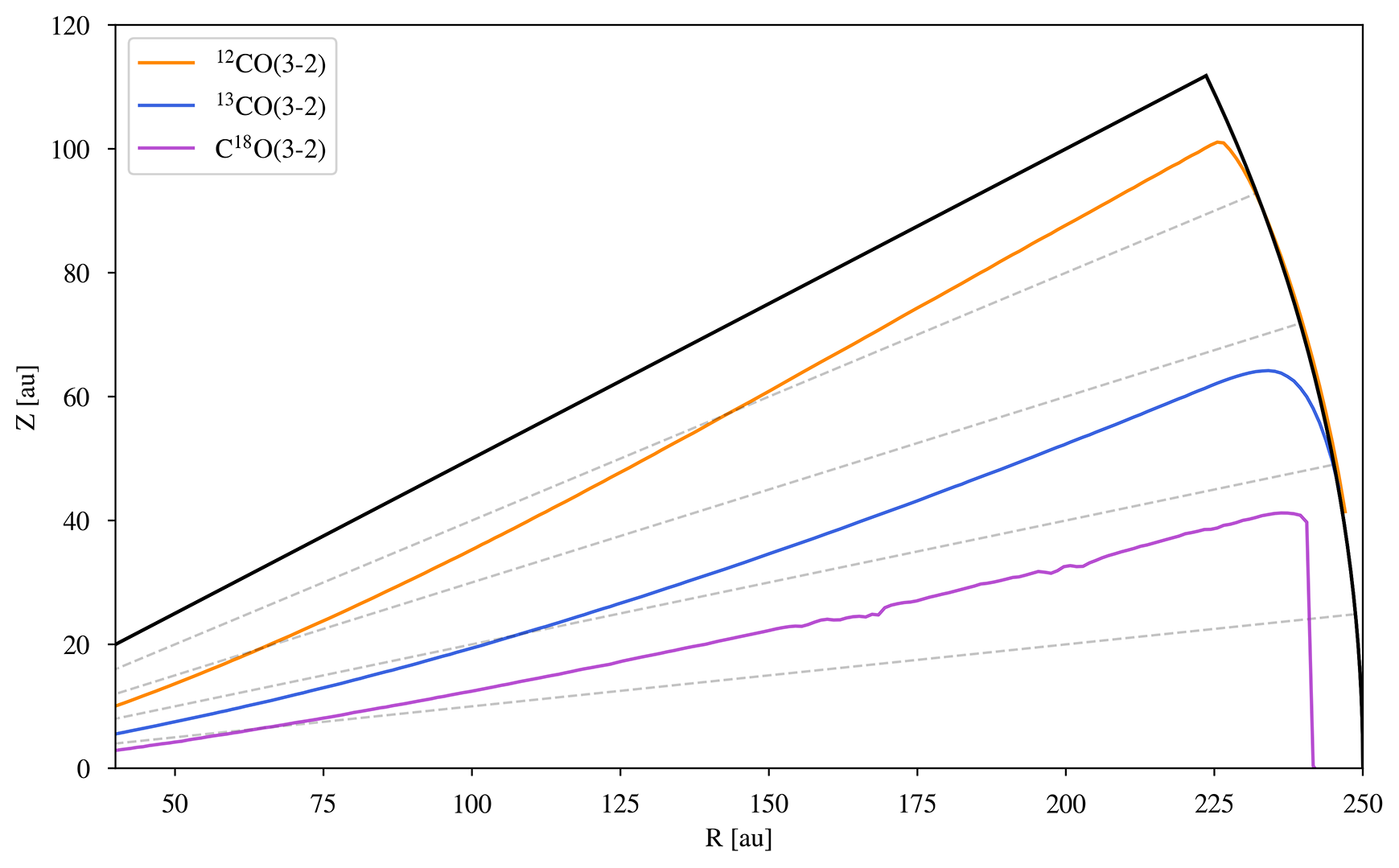}
    \caption{Radial profiles of the surfaces in which the optical depth ($\tau$) reaches one for the J=3-2 transition of $^{12}$CO, $^{12}$CO and C$^{18}$O. The gray dashed lines indicate $Z=\gamma R$ for $\gamma=$0.1,0.2,0.3 and 0.4. The black solid lines indicate the simulation's grid edges ($Z=0.5R$ and $r=250$ au).}
    \label{fig:tau1surfaces}
\end{figure*}

\begin{figure*}[htp]
    \centering
    \includegraphics[width=\textwidth]{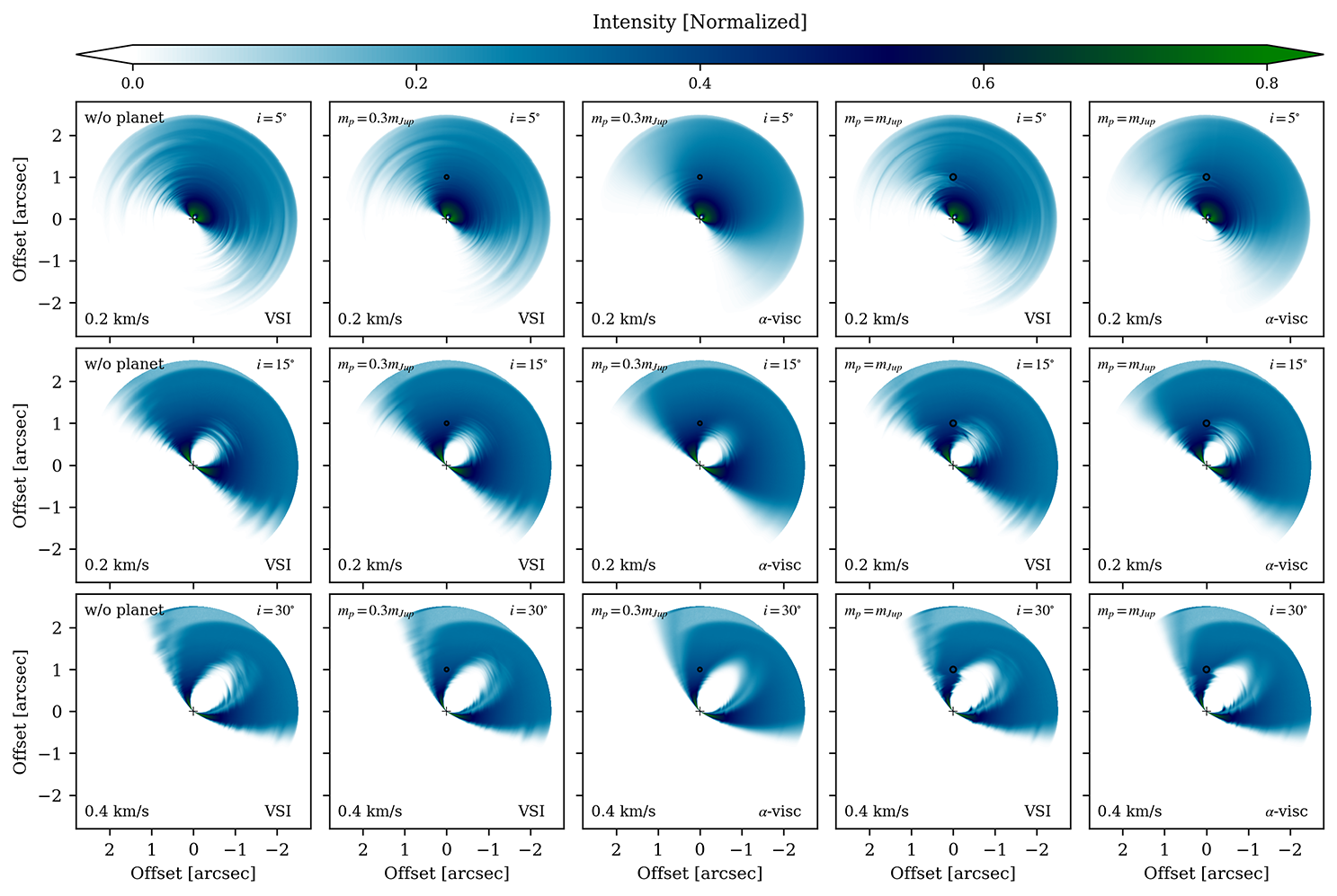}
    \caption{Velocity channel maps for the synthetic $^{12}$CO(3-2) emission line maps. From the left to right columns: VSI unstable disk without a planet, VSI-unstable disk with a Saturn planet, $\alpha$-viscous disk with a Saturn planet, VSI-unstable disk with a Jupiter planet, and $\alpha$-viscous disk with a Jupiter planet. From top to bottom, predictions for different disk inclinations are presented ($i=[5^{\circ}$, $15^{\circ}$, $30^{\circ}]$). The $\alpha$-viscous models are run for $\alpha=5\times 10^{-4}$. The approximated location of the planets is indicated with black circles, of the size of the planet's Hill spheres. Displayed are channels that overlap with the location of the planets. A short movie of all the velocity channel maps is included in the supplementary online material.}
    \label{fig:channelmaps}
\end{figure*}

\begin{figure*}[htp]
    \centering
    \includegraphics[width=\textwidth]{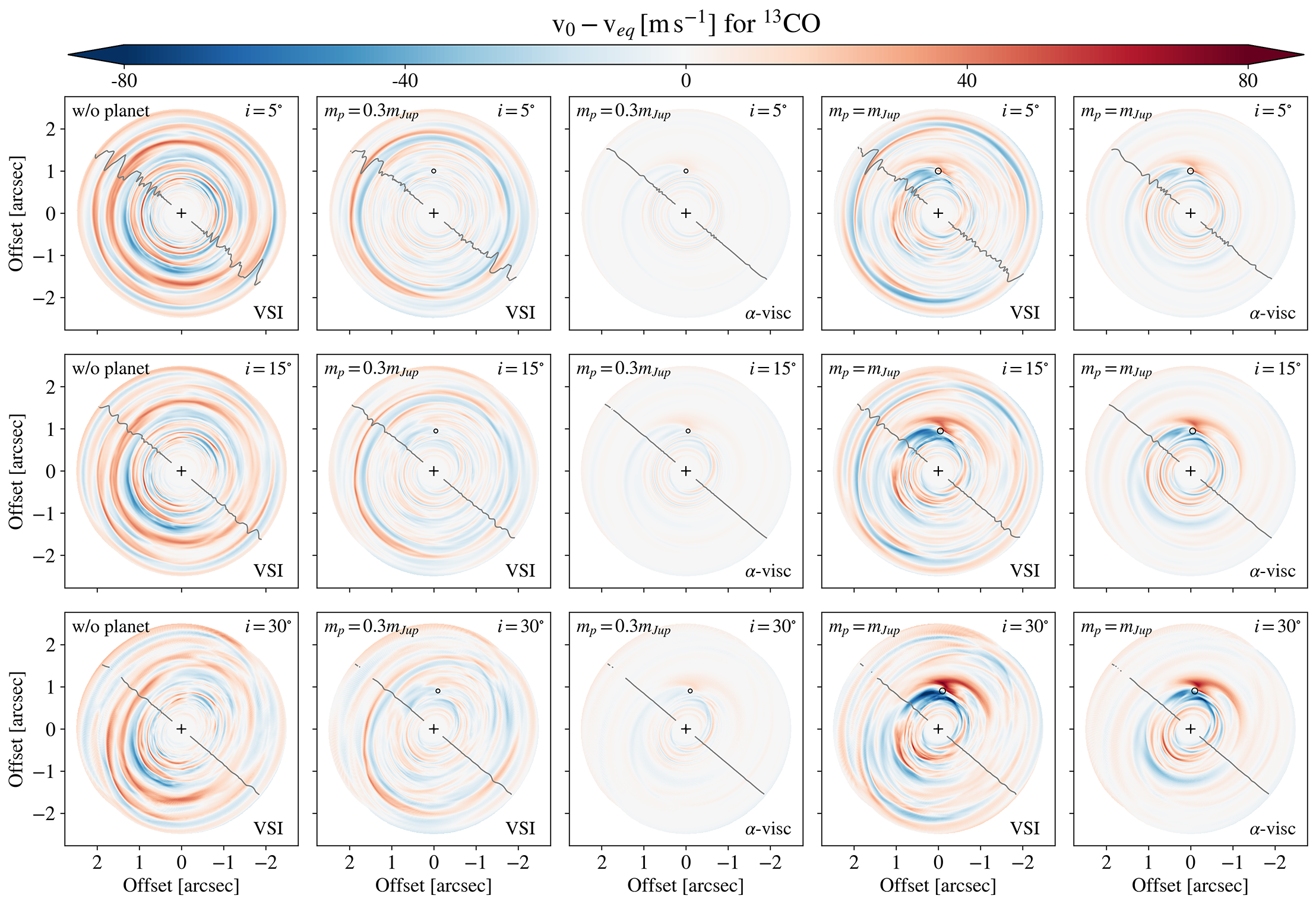}
    \caption{Same as Figure \ref{fig:residualsraw12co} but for the synthetic $^{13}$CO$(3-2)$ predictions.}
    \label{fig:residualsraw13co}
\end{figure*}

\begin{figure*}[htp]
    \centering
    \includegraphics[width=\textwidth]{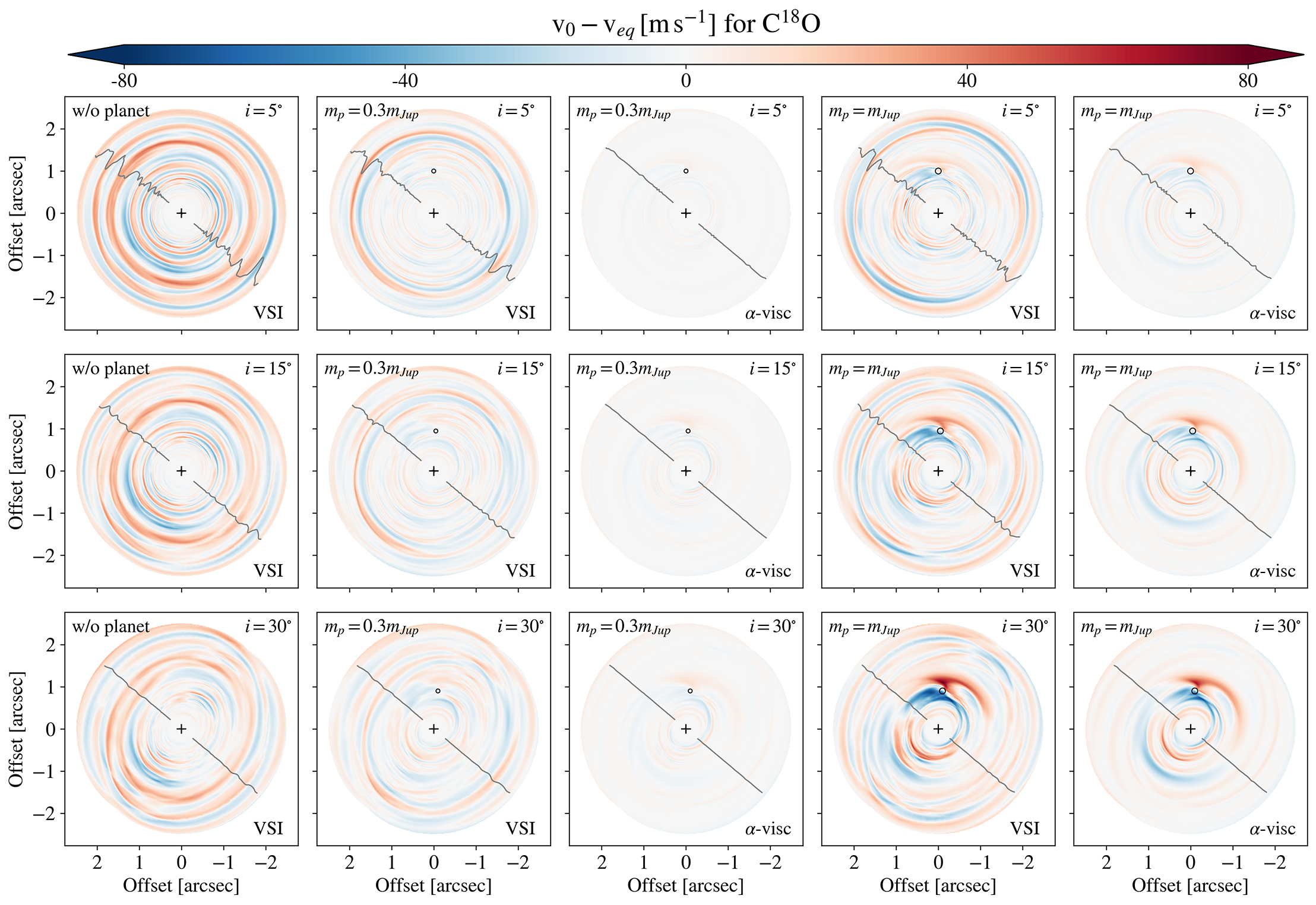}
    \caption{Same as Figure \ref{fig:residualsraw12co} but for the synthetic C$^{18}$O$(3-2)$ predictions.}
    \label{fig:residualsrawc18o}
\end{figure*}

\begin{figure*}[htp]
    \centering
    \includegraphics[width=\textwidth]{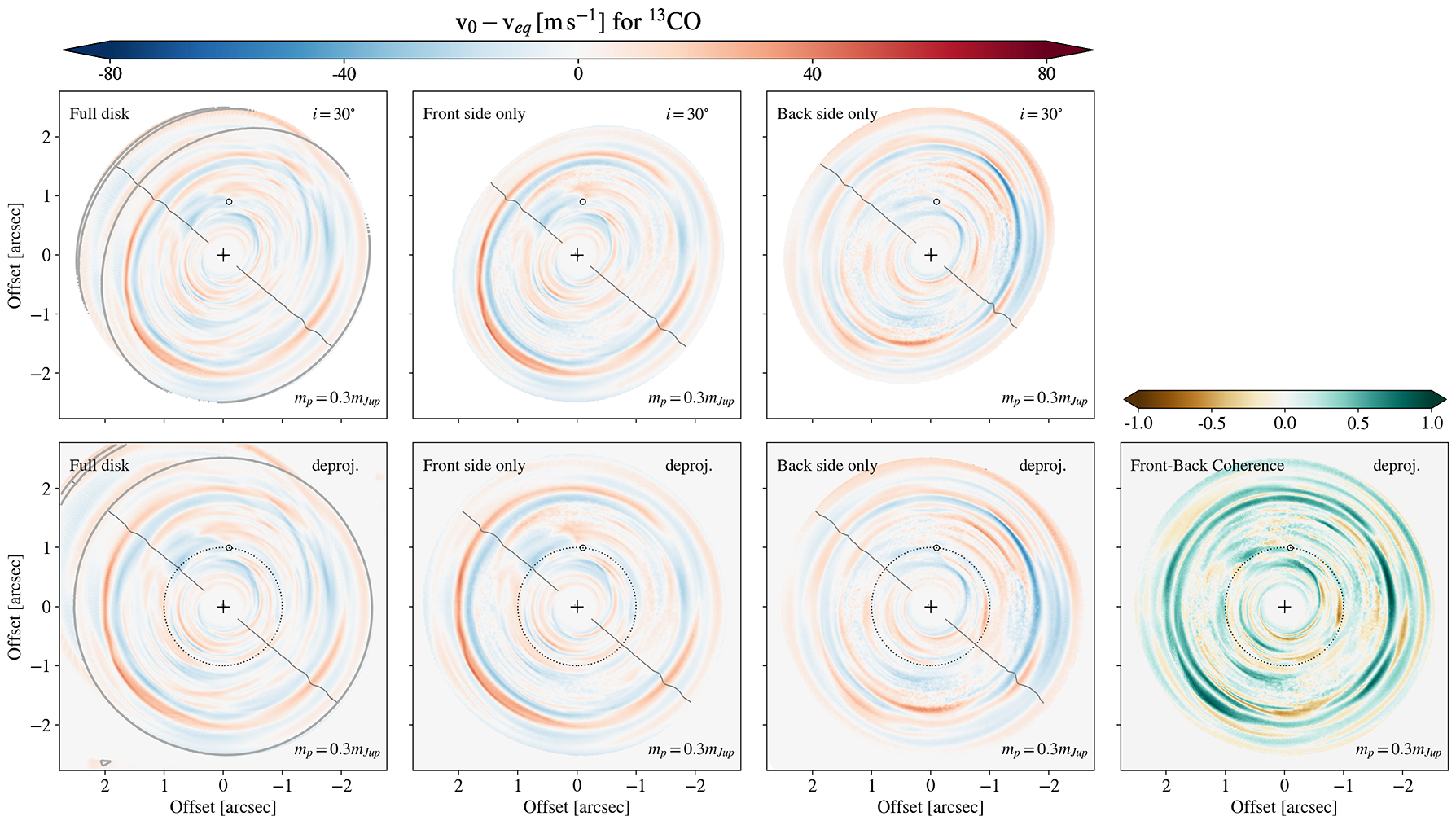}
    \caption{Same as Figure \ref{fig:residualstwolayersvsi}, but for a VSI-unstable disk with an embedded Saturn-mass planet.} 
    \label{fig:residualstwolayerssaturn}
\end{figure*}

\begin{figure*}[htp]
    \centering
    \includegraphics[width=\textwidth]{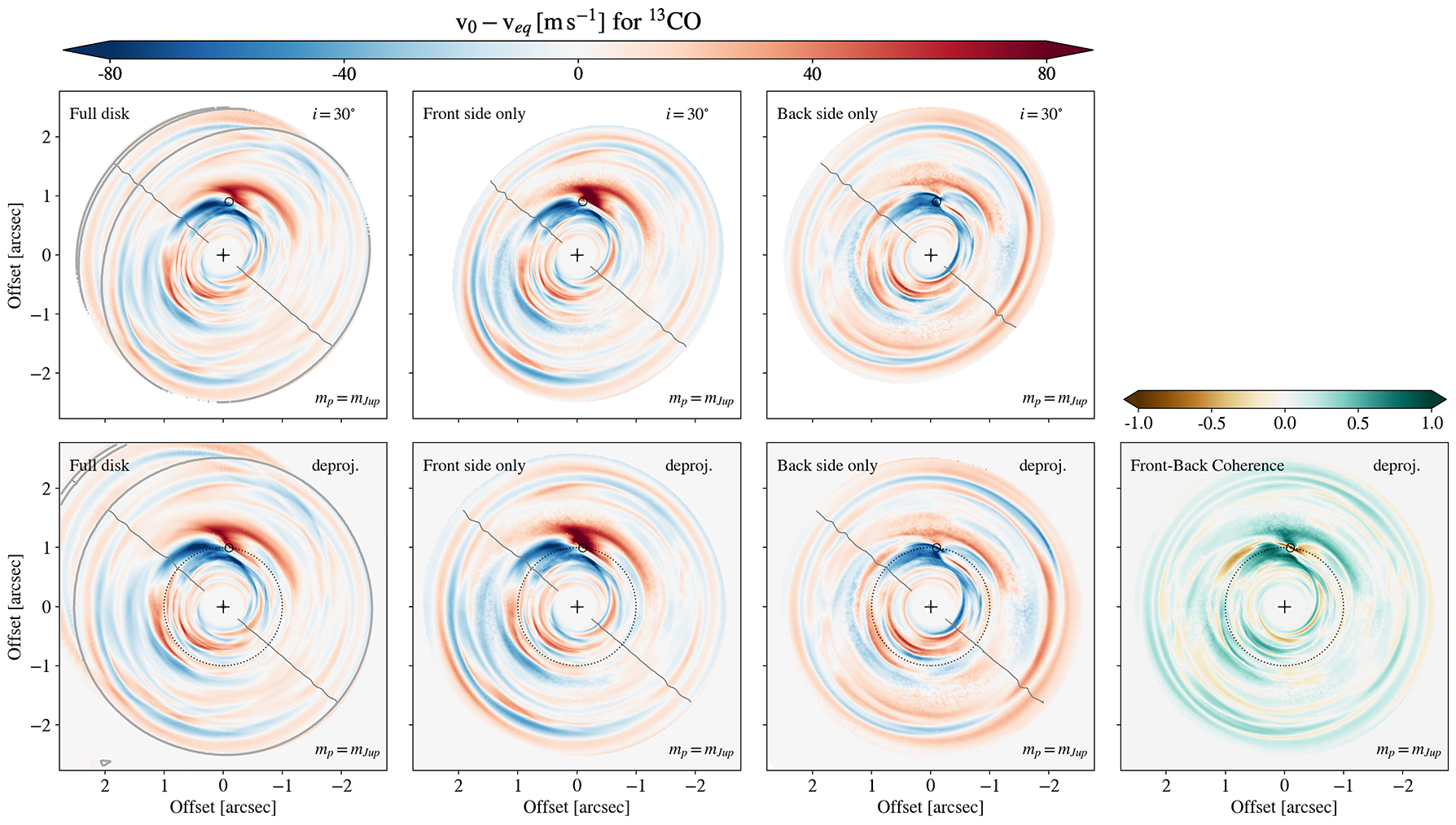}
    \caption{Same as Figure \ref{fig:residualstwolayersvsi}, but for a VSI-unstable disk with an embedded Jupiter-mass planet.} 
    \label{fig:residualstwolayersjupiter}
\end{figure*}

%
%

\end{document}